\documentclass[useAMS,usenatbib]{mn2e}
\usepackage{graphicx}
\usepackage{amsmath}
\usepackage{amsfonts} 
\usepackage{color}
\title[Collisional parameters of planetesimal belts
]
{Collisional parameters of planetesimal belts, precursor of 
debris disks, perturbed by
a nearby giant planet.}

\author[]{F. Marzari$^{1}$ and A. Dell'Oro$^{2}$ \\
$^{1}$Dept. of Physics, University of Padova, 35131 Italy\\
$^{2}$INAF, Osservatorio Astrofisico di Arcetri, Largo E. Fermi 5, I-50125, Firenze, Italy}
\begin{document}

\date{Accepted .....;  Received ..... ; in original form ........}

\pagerange{\pageref{firstpage}--\pageref{lastpage}} \pubyear{.....}

\maketitle

\label{firstpage}

\begin{abstract}

Planetesimal belts are invoked to explain the prolonged 
existence of debris disks.   
Important parameters to  
model their collisional evolution and to 
compute the dust production rate are the intrinsic 
probability of collision $P_i$ and the mean impact velocity $U_c$. 
If a planet orbits close to the belt, the values of both these parameters
are  
affected by its secular perturbations yielding a strong correlation 
between eccentricity $e$ and pericentre longitude $\varpi$.  
We adopt a new algorithm to compute 
both $P_i$ and $U_c$ in presence 
of various levels of secular correlation   
due to different ratios 
between proper and forced eccentricity. 
We tested this algorithm in a 
standard case with a Jupiter--sized planet 
orbiting inside a putative 
planetesimal belt finding that it
is 
less collisionally active compared to a 
self--stirred belt because of the $e - \varpi$ coupling. 
The eccentricity of the planet is an important parameter 
in determining the amount of dust production  since
the erosion rate is 10 times faster when the planet eccentricity 
increases from 0.1 to 0.6. Also the initial conditions of the belt 
(either warm or cold) and its average inclination 
strongly affects  $P_i$ and $U_c$ and then 
its long term collisional evolution in
presence of the planet.
We finally apply our method to the planetesimal belts
supposedly refilling the dust disks around HD 38529 and $\epsilon$ Eridani. In the 
most collisionally active configurations, only a small 
fraction of bodies smaller than 100 km are expected to be 
fragmented over a time--span of 4 Gyr. 
\end{abstract}


\section{Introduction}

Debris disks, mostly detected in thermal infrared, are observed 
around 10--20\% of solar type stars \citep{hillenbrand2008,
trilling2008, sibthorpe2013}. The micron--sized grains populating 
these disks are short lived mostly because of radiation related forces
and collisional erosion. As a consequence, their refilling requires the 
presence of a reservoir of planetesimals whose collisions 
constantly produce new dusty debris. 
Indeed debris disks are the only signature of these belts whose 
properties like size distribution, mechanical 
strength and dynamical excitation cannot be constrained by 
observations. The only way to link planetesimals and dust is 
through numerical models reproducing the SED and resolved images 
of debris disks. However, these models not always give 
unique solutions. 

Analogues of the solar system 
asteroid or Kuiper belt, planetesimal belts are leftovers from the 
planet formation process and they eventually continue to evolve
under mutual collisions. 
The frequency of debris disks as well as the infrared excess strength 
decline with the stellar age \citep{krivov2010} 
suggesting that the collisional process slowly grinds down the 
initial planetesimal population. Different mechanisms have been 
invoked to stir up a planetesimal swarm igniting mutual destructive 
collisions like self--stirring by 
larger planetesimals \citep{chambers2001,kenyon2004} and
perturbations by planets \citep{mustill2009}.
Concerning this last mechanism, it is still controversial 
from an observational point of view 
if there is any significant correlation between stars 
with dust emission and the presence of known planets \citep{bryden2009,moro2007b}
even if some of such systems are already known like that in $\epsilon$ 
Eridani \citep{benedict2006}, Fomalhaut \citep{kalas2008} and 
HD 38529 \citep{moro2007}. From a
theoretical point of view,
it is expected that debris disks and planets coexist in 
a large number of systems  being both the outcome of dust coagulation 
and planetesimal accumulation even if 
debris disks appear to be more common than massive planets \citep{moro2007b}.
This may be due to the presence  of debris disks even in systems where the planetesimals 
were not able to form the core of giant planets. In this scenario, the 
dust would be produced in mutual collisions possibly triggered by self--stirring. 
There are significant observational problems in statistically assessing the 
probability of finding giant planets and debris disks coexisting in the 
same system. Young stars have a higher percentage of debris disks but 
radial velocity surveys for planets around them are difficult because of the 
noise due to stellar activity. On the other hand, planets can be easily detected 
around old stars where debris disks may have been eroded away by collisional 
evolution and are expected to be less common. It is also difficult to precisely locate the debris 
belts in radial distance in order to asses the potential detectability of 
close-by planets. 
Finally, it should be noted that the inventory of giant exoplanets is still 
poor for semimajor axes larger than a few AU. 

Assuming that in some systems different dynamical mechanisms lead to a 
configuration with a massive planet 
in an inside orbit respect to a debris disk, this would be a particularly interesting
architecture
since the disk structure is very sensitive to the planet 
gravitational perturbations  
potentially driving the formation of 
arcs, gaps, warps and asymmetric clumps in the disk \citep{moro2005,moro2007}.
Even the gap between two components debris disks is suspected 
to be carved by intervening planets scattering away the remnant planetesimals \citep{su14,
shannon16}
and a survey based on direct imaging is searching for planets 
in systems with double debris disks \citep{mesh15}.
In addition to these direct effects,  a planet also perturbs 
the planetesimal belt stirring up their orbits and affecting the 
collision probability and impact velocity \citep{mustill2009}. The 
level of stirring depends on the planet mass, vicinity 
to the belt, either inside or outside, and orbital eccentricity. 
An excited planetesimal belt may give origin to a brighter debris 
disk at start \citep{wyatt07}, but its lifetime will be significantly shorter 
due to the collisional erosion. 

We can envisage different scenarios 
related to the coexistence of planets and debris disks.
A single giant planet, which had a limited amount of inward 
migration, can clear part of the leftover planetesimal population leaving an
inner cavity in the planetesimal disk. In this case we would find a low 
eccentricity planet secularly perturbing an external planetesimal belt. 
This configuration does not exclude the presence of additional planets 
in the system moving in inside orbits. 
More complex dynamical configurations are expected when a multi--planet system
undergoes a period of extended chaotic evolution 
characterised by planet--planet scattering. In extreme cases this 
dynamical evolution may lead to a complete clearing of local 
planetesimal belts because of the  
intense perturbations of the 
planets on highly eccentric orbits \citep{bonsor2013,marzari2014}. 
However, less violent evolutions may lead to a final architecture where a giant planet
on an eccentric orbit
is located inside, or outside, the surviving belt and it affects its evolution via secular
perturbations. The two different mechanisms may lead to a wide range of possible 
eccentricities for  
the planet perturbing the belt. 

When a planetesimal belt is accompanied by a close-by planet,
the relative impact 
velocity and frequency of collisions of the perturbed belt, used to 
predict its long term collisional evolution, cannot
be calculated with the {\"O}pik/Wetherill analytic formulation
\citep{opik51,wetherill67} even in the improved formulations developed 
to study the evolution of the asteroid belt (see 
\cite{davis02} for a review). In presence of a massive 
planet, in particular if on a highly eccentric orbit,  
the forced component in the secular evolution of the 
planetesimal orbits may lead to a strong correlation between  
eccentricity and perihelion longitude. This correlation 
invalidates the  {\"O}pik/Wetherill methods based on a uniform 
distribution of the orbital angles (perihelion longitude 
and node longitude) derived under the assumption of periodic circulation of 
the two angles. 
These methods work properly  when describing the evolution of the 
asteroid or Kuiper belt in the solar system where the 
forced eccentricity is significantly smaller than the proper one, 
but in dynamical configurations where the planetesimal ring is perturbed 
by a giant planet on a highly eccentric orbit they fail.

A first attempt to overcome this problem is described in 
\cite{mustill2009} where they first derive refined values of the 
forced eccentricity from which they 
compute an average value of the impact velocity between the 
planetesimals multiplying the forced eccentricity by the local
Keplerian velocity and a constant value $c \sim 1.4$. 
In this paper we apply an innovative semi-analytic method to estimate 
both the intrinsic probability of collision $P_i$  i.e. the collision rate
per unit cross--section area of target and projectile per unit time, 
and the average collision speed $U_c$, derived from a detailed 
statistical frequency distribution of the relative velocities,  in a planetesimal belt perturbed 
by an eccentric giant planet. This method is designed in order to  
fully account for the  different levels of correlation 
between eccentricity and perihelion longitude of the planetesimals 
caused by the secular dynamics. 
Both $P_i$ and $U_c$  are needed to properly model the collisional evolution
of a planetesimal disk and predict the  
dust production rate refilling the related debris disk. In particular,
$P_i$ is more relevant in establishing the amount of erosion of the belt.  
We consider two distinct plausible scenarios where, prior to the 
evolution of the perturbing planet close to the belt, the 
planetesimal disk was either dynamically non--excited (cold population) or 
excited (warm population). In the former case (cold population), when the planet 
approaches the belt during its evolution, the proper eccentricity of the planetesimals is 
immediately excited to a value approximately equal to the secular forced one 
\citep{thebault2006}. This is also the scenario explored by \cite{mustill2009} 
and it is the dynamical configuration where the correlation between 
$e$ and $\varpi$ is maximised. If instead 
the eccentricities of the planetesimal
population were already significantly excited when the planet sets in and begins to 
perturb the belt (warm belt) a 
different dynamical configuration is achieved where the proper and forced 
eccentricities  may significantly differ.  We will 
consider cases where 
the eccentricity, prior to the onset of the secular perturbations, 
is an increasing fraction of the forced eccentricity. In these cases
after the onset of the secular perturbations 
the degree of correlation between $e$ and $\varpi$ will be less robust 
within the belt and pseudo--librator states will appear in the population
influencing the values of both $P_i$ and $U_c$. 
While modelling a warm belt, it is reasonable to expect that also the 
inclination is  excited as well so we will 
explore the effect of a high planetesimal inclination on both 
the collisional parameters. 

If a belt is densely populated by planetesimals, a significant collisional damping of the 
eccentricity may 
occur \citep{damp88} leading it to a state of cold belt before the planet approaches. After the 
planet sets into a perturbing orbit, the collisional damping may  
still be efficient in reducing the eccentricity, but in this 
case it will affect only the proper term causing its progressive decrease. We explore 
also this scenario and estimate the values of the collisional parameters to 
compare with non--damped cold and warm belts.

In Section 2 we briefly summarise the expected dynamical behaviour of 
a planetesimal belt perturbed by a planet. 
In Section 3 we 
outline the method developed to compute the average values of 
intrinsic probability of collision $P_i$
and impact velocity $U_c$ for a belt characterised by the 
above mentioned secular 
dynamics. In Section 4  we apply the method to a 
'standard' case  to illustrate the effects of the secular dynamics 
and compare the predicted values of 
$U_c$ with those derived by \cite{mustill2009}.
We also derive and compare the values of the collisional parameters in 
cold and warm belts with and without inclination excitation. 
In Section 5 we model the collisional evolution of a putative 
belt with the previously estimated collisional parameters 
while in Section 6
we model a real system, HD 38529, and compare its evolution 
with that of $\epsilon$ Eridani.
Finally, in 
Section 7 we summarise and discuss our results. 

\section{Secular evolution and correlation between eccentricity and perihelion longitude}

The dynamical evolution of a minor body population perturbed by a
planet is classically described by the linear secular theory of 
Laplace--Lagrange \citep{murray1999}. For a mass--less planetesimal population, 
the evolution with time of 
the non--singular variables $h = e\sin \tilde \omega$ and $k = e \cos \tilde \omega$, where
$e$ and $\tilde \omega$ are the eccentricity and longitude of the pericentre of the
osculating orbit, respectively,  is 
given by 

\begin{equation}
\begin{aligned}
\label{eq:secular}
h &= e_p sin (A t + B) + e_f \\ 
k &= e_p cos (A t + B)  
\end{aligned}
\end{equation}

where $B$ is a constant determined by the initial conditions of the 
system,  $e_p$ is termed proper eccentricity while $e_f$ is the forced one.
In our case we consider the simplified situation in which the planet is 
not perturbed by additional bodies so that $e_f$ is constant.
In Eq. \ref{eq:secular} we assume that the reference frame for the 
computation of the orbital elements is aligned with the apsidal line 
of the planet orbit. Approximate values for the proper frequency $A$ and 
the forced eccentricity $e_f$ can be derived from the simplified linear
disturbing function \citep{murray1999} and are given in \cite{mustill2009}

\begin{equation}
\begin{aligned}
e_f & \sim {5 \over 4}  \alpha e_{pl} \\ 
A & \sim n {3 \over 4} {m_{pl} \over  m_s}  \alpha^2 \bar{\alpha}
\end{aligned}
\label{ef}
\end{equation}

where $m_s$ is the mass of the central body (the star), 
$e_{pl}$ the eccentricity of the planet and $n$ is the mean 
motion of the test body. For a configuration in which the planet is interior 
to the planetesimal orbit  
$ \alpha = {a_{pl} \over a}$ and $ \bar{\alpha} = 1$ 
while 
$ \alpha = {a \over a_{pl}}$ and $ \bar{\alpha} = \alpha$ for an exterior planet.
$a$ is the 
semimajor axis of the planetesimal and $a_{pl}$ the  semimajor axis 
of the planet \citep{murray1999,mustill2009}. This classical perturbation theory works far from 
mean motion and secular resonances and, as a consequence,  some values of $ \alpha$ lead to 
incorrect predictions. In addition, 
the theory has been developed to second order 
in the eccentricity and inclination of the bodies and it is
a good approximation only for small values of these orbital parameters. 
Among known extrasolar planets, a significant fraction have high orbital eccentricities 
for which 
either 
semi--numerical approaches \citep{micht2004} or higher order theories 
\citep{libert2005} are preferable.   However,  a
numerical exploration 
of the reliability of the formulas of the linear theory even for large values of $e_{pl}$ by
\cite{mustill2009} has shown that the precession rate can be well 
described by the disturbing function developed by \cite{heppenheimer1978}
for binary stars:

\begin{equation}
A \sim 2 \pi n {3 \over 4}  {m_{pl} \over m_s}  {{\alpha^2 \bar{\alpha}} 
\over {4 (1 - e_{pl}^2)^{3/2}}} 
\label{A2}
\end{equation}

\noindent
for values of $e_{pl}$ beyond 0.2 even if the correction proposed by \cite{thebault2006} was not tested.
This last is a more accurate prescription for the frequency of the secular 
oscillations induced by a binary companion empirically derived from several 
numerical simulations. In our context, a very precise value of the 
frequency A is not strictly required since what matters is that the 
randomisation of the angles is achieved.  
A constant frequency of circulation 
is the only assumption about $\theta = At+B$ used
in our statistical model.
The forced eccentricity $e_f$ is instead 
well approximated by the linear secular theory even for large values of $e_{pl}$. 

\begin{figure}
  \includegraphics[width=\hsize]{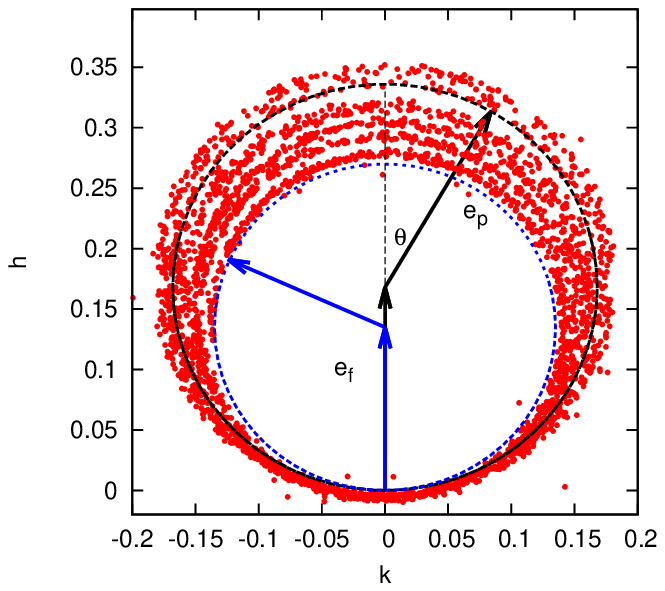}
  \includegraphics[width=\hsize]{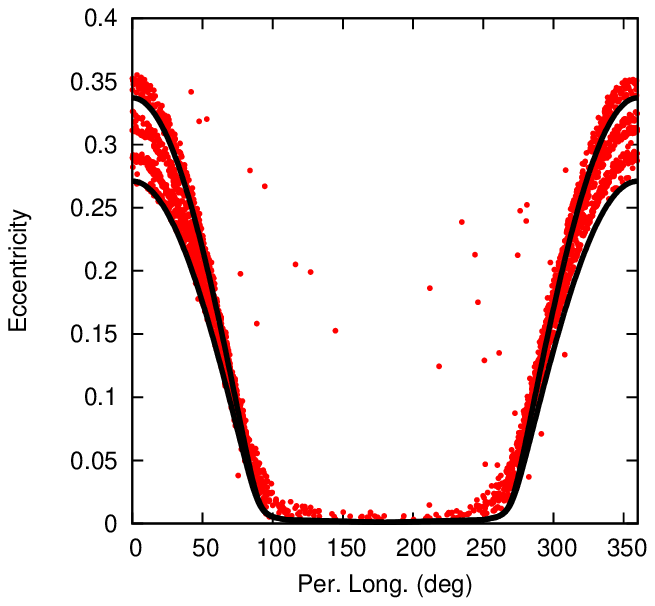}
  \caption{\label{secular}
Secular evolution of a population of planetesimals moving in a ring with
semimajor axis ranging from 15 to 20 AU and initially circular 
orbits. They are perturbed by a Jupiter--size planet with $a_{pl} = 5$ AU
and eccentricity $e_{pl}=0.5$. In the top panel the evolution of the 
non--singular $h$ and $k$ variables is illustrated ($h = e \ cos(\varpi)$, $k = e \ sin(\varpi)$). 
The arrows show the 
forced $e_f$ and proper $e_p$ eccentricities for $a = 15$ AU (in blue)  and 
$a = 20$ AU (in black), which, once combined, give
the eccentricity of the planetesimals. The blue and black dashed lines show 
the curves obtained by combining proper and forced eccentricity.
On the bottom panel the 
correlation between $\varpi$ and $e$ is highlighted. 
}
\end{figure}

To test the reliability of the linear secular theory in dealing with the main dynamical features 
of a planetesimal belt perturbed by an eccentric planet, we have compared the outcome of a short term numerical 
simulation with the predictions of the theory. 
In Fig.\ref{secular} we show the evolution with time of a population of planetesimals
orbiting between 15 and 20 AU on initially circular orbits and perturbed 
by a planet at 5 AU with an eccentricity $e_{pl} = 0.5$. 
The 15th order
RADAU integrator \citep{Everhart85} has been used for this short term simulation. 
In the upper
panel we show the $h$ and $k$ variables after about 10 Myr of evolution.
The pericentre longitude is randomised since the period of 
the secular circulation is lower than 1 Myr and the integration time--span is also significantly 
longer than the 'crossing time' defined in \cite{mustill2009} 
for an internal perturber. This time is
given by: 

\begin{align}
 t_{cross} &  \sim 1.53 \times 10^3 {{(1 - e_{pl}^2)^{3/2}} \over {e_{pl}} }
\left ( {a \over {10 AU}} \right )^{9 \over 2} \times \nonumber  \\  
& \left ( {M_* \over M_{\odot} } \right )^{1 \over 2} \times 
\left ( {m_{pl} \over M_{\odot}} \right )^{-1} \times
\left ( a_{pl} \over {1AU} \right )^{-3} yr,
\end{align}

\noindent
and, for our test belt, the longest $t_{cross}$ is about $3.6 \times 10^5$ yr. 

The theoretical evolution for $e$ and $\varpi$  predicted by the linear secular theory
can be derived combining the two vectors drawn in the plot 
representing the forced $\bf e_f$ and proper $\bf e_p$  eccentricities. In the 
bottom panel the planetesimal eccentricity is shown as a function of 
the perihelion longitude $\varpi$  and compared to the theoretical curve. 
The analytic curves are computed for a single value of semimajor axis
and, as a consequence, they only match the evolution of the planetesimals 
with initial 
semimajor axis similar to that used in the computation of the curves. 
The agreement between the linear theory and the numerical results is really good apart from a 
few scattered points which are due to the effects of mean motion resonances
with the planet and can be neglected. 

\begin{figure}
  \includegraphics[width=7truecm,height=6truecm]{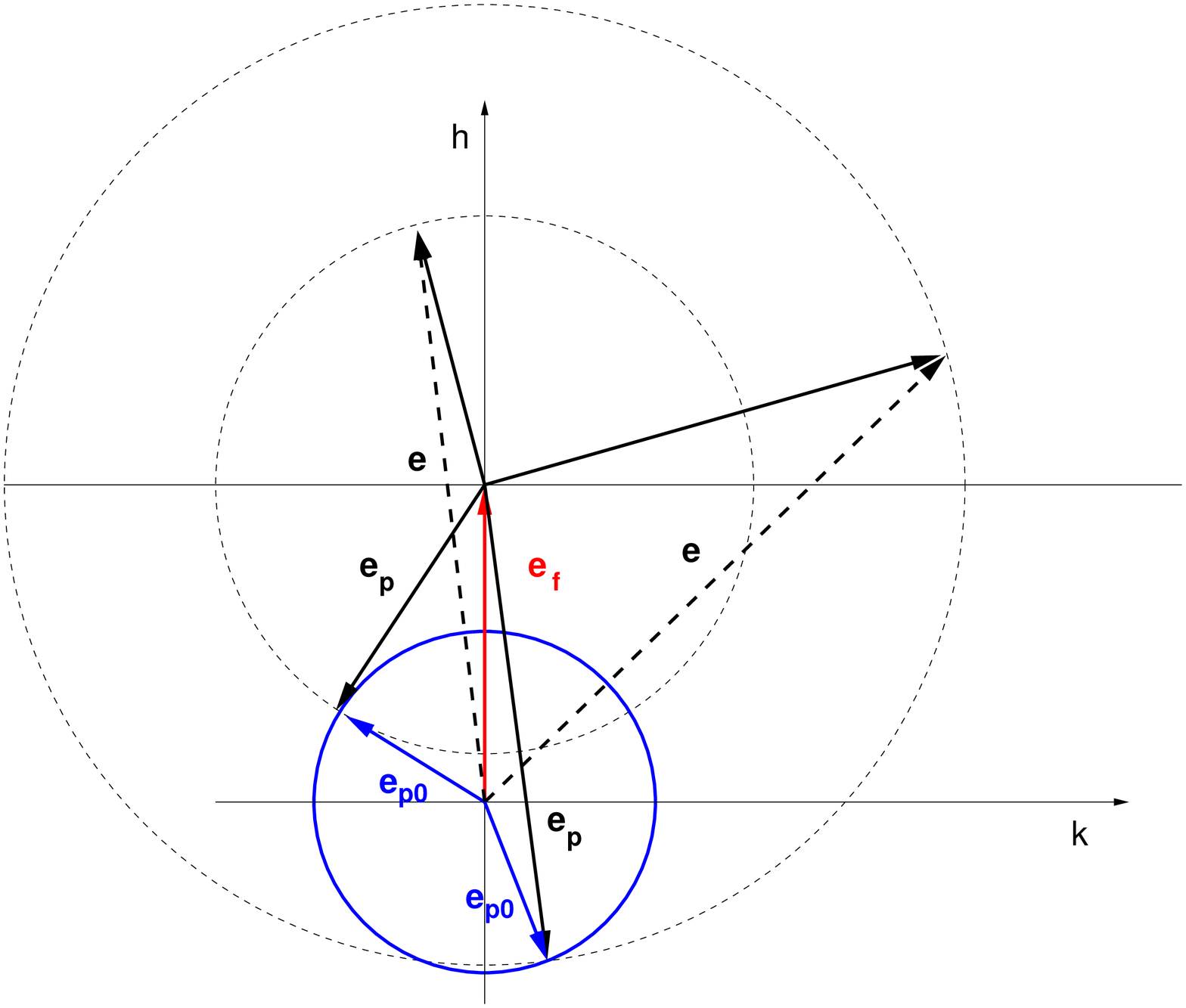}
  \includegraphics[width=\hsize]{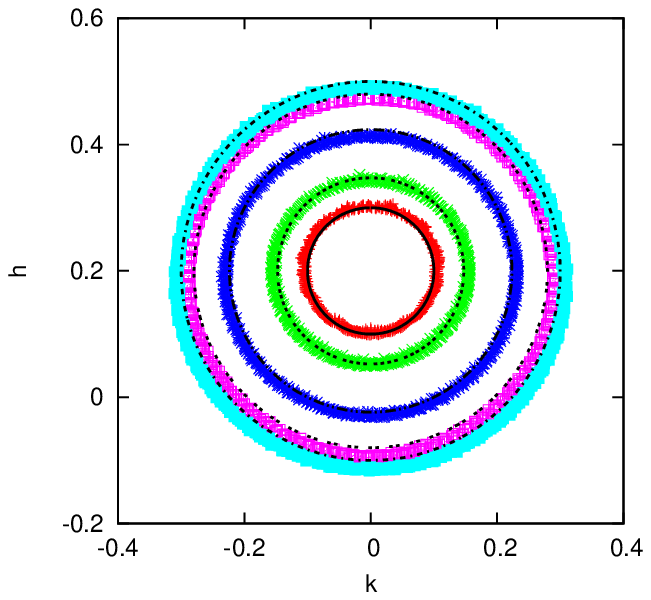}
  \includegraphics[width=\hsize]{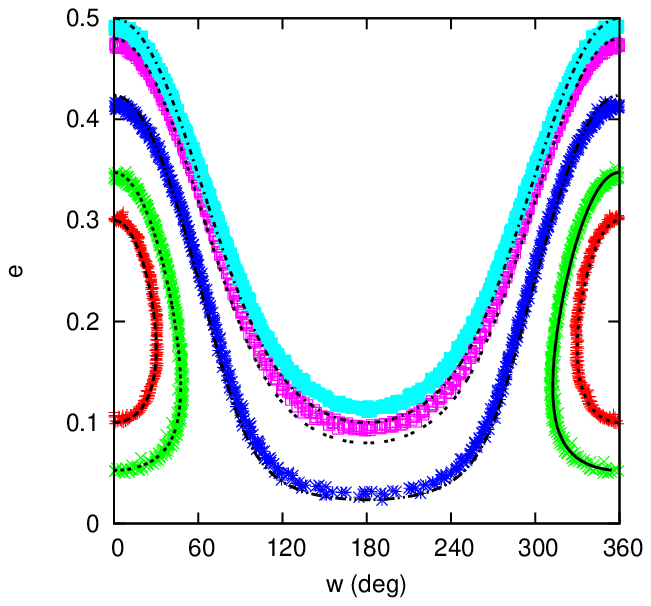}
  \caption{\label{secular2}
In the upper panel the secular dynamics is sketched. The blue circle describes the location of the
tip of the proper eccentricity vector $\bf e_{p0}$  in an unperturbed warm belt.
When the planet perturbs the belt, new initial proper vectors $\bf e_p$ are computed
joining the tip of $\bf e_f$ with the tip of $\bf e_{p0}$. Once the initial $\bf e_p$s are computed in
this way, the secular evolution can be predicted following the circle centred on the tip of $\bf e_f$ and
depicted by the tip of $\bf e_p$.  In the middle and lower panels the secular behaviour is compared to the outcome of
numerical integrations. We select 5 orbits with semimajor axis equal to 15 AU,
initial values $e_{p0} = 0.5 \cdot e_f $ and 5 different
values of the initial pericentre longitude $0^o$, $45^o$, $90^o$, $135^o$, and $180^o$.  The analytic predictions
are given by the dashed continuous lines and they match closely the numerical results drawn by different colours
depending on the initial value of $\varpi$.  In both panels, for small values of $e_p$, pseudo--librations
around $0^o$
are observed (red and green points).
}
\end{figure}

In the figures, we have assumed that $e_p \sim e_f$, condition which 
is based on the premise that, 
before being perturbed by the planet, the 
planetesimals were initially on 
unperturbed almost circular orbits with $e_{p0}  \sim 0$.  
The approximation $e_p \sim e_f$ 
is common
when exploring the perturbations of a massive 
body on an initially cold  planetesimal belt \citep{heppenheimer1978, whitmire1998}.
In particular, if the planet is injected on a highly eccentric
orbit after a period of dynamical instability, the belt of 
planetesimals will suddenly feel the strong secular perturbations 
of the planet. If we assume that as 'time 0' of the secular 
perturbations the  belt was cold with $e_{p0}  \sim 0$, then 
the condition $e_p \sim e_f$ is naturally imposed in the 
subsequent evolution. 
We term here $e_{p0}$ the proper eccentricity before the onset of the planet 
perturbations. In this case it would not be correct to call it a 'proper' eccentricity
since it is just the average eccentricity of the cold belt. However, since it will determine
the subsequent value of the proper eccentricity once the planet 
begins to perturb the belt, hereinafter we will use $e_{p0}$ to indicate it. 

A scenario different from that of an initial cold belt is 
produced by a belt which has been excited before the 
arrival of the perturbing planet, a warm belt. Various mechanisms 
may contribute to the dynamical excitation like 
stellar flybys, self--stirring and planet stirring related to an
extended period of planet--planet scattering (see \cite{matthews14}
for a review). This last mechanism, due to its large variety of possible
outcomes, may activate the 
planetesimal belt inducing high eccentricities and inclinations 
before a planet is deposited on a finally stable orbit close to 
the belt \citep{bonsor2013,marzari2014}.  In a warm belt 
the initial planetesimal
eccentricity $e_{p0}$ is significantly higher than 0 at the time of the 
planet approach. The secular modelling of an initially warm belt is 
more complex respect to that of a cold belt since in the former case the initial 
conditions for the secular evolution are different from the simple assumption 
$e_p \sim e_f$. In  Fig.\ref{secular2} we show how 
the new proper eccentricity $e_p$ of  each planetesimal in the belt,
after the planet perturbations are switched on, can be computed  from  $e_{p0}$ 
(initial average eccentricity of the belt before the approach of the planet)
and how the secular theory can be used to predict the subsequent dynamical evolution. 
In the upper panel of Fig.\ref{secular2} we sketch the mode of operation of the secular 
theory  
using vector formalism.  
The blue circle marks the location of the tips of all the initial
proper eccentricity vectors
$\bf e_{p0}$'s of the warm belt. All these vectors have the same  modulus 
(we assume that all planetesimals have the same initial $e_{p0}$ equal to the 
average of the belt)
while their orientation is random depending on $\varpi$.
When the planet begins to perturb the belt, a common forced eccentricity vector $\bf e_f$
appears in the secular evolution.  As a consequence, each body will acquire a new 
proper eccentricity vector ${\bf e_p}$ which depends on the 
the initial proper eccentricity vector ${\bf e}_{p0}$ through the relation
${\bf e}_f + {\bf e}_p = {\bf e}_{p0}$. The new 
proper eccentricity 
vector ${\bf e}_p = {\bf e}_{p0} - {\bf e}_f$ will 
mark the secular evolution of each planetesimal from then on. 
The modulus of the new ${\bf e_{p}}$ ranges from $|e_f - e_{p0}|$ and  $e_f + e_{p0}$ with 
all intermediate values being possible, if the belt is enough crowded. 
The $ \bf e_{p}$'s will be used, together with $\bf e_f$, to compute the 
subsequent secular evolution featured by circles centred on the tip of $\bf e_f$ and 
with radius equal to $e_{p}$.  In the middle and lower panels of Fig.\ref{secular2} 
we compare the analytic predictions of the secular theory with numerical integration of five different 
planetesimal orbits. They are all started with semimajor axis $a = 15$ AU  and 
all have the same value of  
initial eccentricity $e_{p0} =0.5 \cdot  e_f $. The
pericentre longitude $\varpi$ is set to $0^o$, $45^o$, $90^o$, $135^o$, and $180^o$,
respectively. In this way we cover the most relevant secular behaviours typical of an initially
warm belt perturbed by a planet. For small values of $e_p$ a pseudo--libration around $0^o$ is observed 
while for larger values circulation is restored. In all cases there  is a significant 
correlation between eccentricity and pericentre longitude. 

In modelling warm belts we will sample 
three different initial ratios between 
$e_{p0}$ and $e_f$ i.e. 0.25, 0.5, and 1. Linking $e_{p0}$ to $e_f$  is an arbitrary choice but it is adopted to 
give an idea of how higher initial values of $e_{p0}$ influence both $P_i$ and $U_c$ and it
appears more robust than selecting random values of $e_{p0}$. It is less
arbitrary, it can be easily replicated and, in addition, it includes a radial dependence 
which may be present in the initial belt. At present we do not have the means to
reconstruct the past history of a warm belt and of the dynamical mechanisms which may
have stirred it (planet formation, protoplanets roaming around, a phase of planet-planet scattering
etc...) and, as a consequence,  many different initial distributions of the planetesimal proper eccentricities 
are conceivable.  We focus in this paper on those where the proper eccentricity does not 
exceed the forced one and decreases with radial distance. 
For 
warm belts we will explore a scenario where the dynamical excitation involves also the 
inclination
modelling, together with the case with $i=3^o$,  one with 
$i= 15^o$.

\begin{figure}
  \includegraphics[width=\hsize]{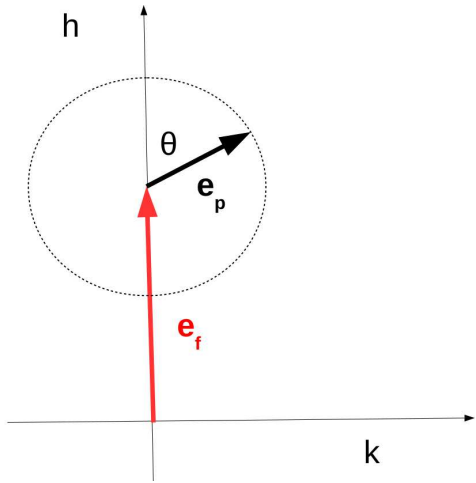}
  \caption{\label{secular3} If the planetesimal belt is densely populated, 
a significant collisional damping may occur reducing the proper eccentricity 
even after the planet ignited its secular perturbations. In this case the 
secular evolution will lead to pseudo--librations of the pericentre of all
bodies and the secular evolution is described by the dashed circle. 
}
\end{figure}

If the planetesimal belt is densely populated, a strong collisional damping may occur
even after the planet reached its close-by orbit. This damping is due to the loss of 
orbital energy after each collision and, on average, it causes a reduction of the 
proper eccentricity $e_p$. Any crowded belt might be drag to a condition 
similar to that illustrated in Fig. \ref{secular3} even in presence of a
perturbing planet. The proper eccentricity is smaller than 
the forced one and all planetesimals evolve in pseudo--librating orbits with a degree 
of alignment which depends on the ratio between proper and forced eccentricity. 
We will model also this dynamical configuration and compute values of $P_i$ and $U_c$ in 
this scenario. 

We have to point out that our estimates of $P_i$ and 
$U_c$ cannot be applied to very young belts, since our model requires 
a full randomisation of the pericentre of the planetesimals. 
This occurs on a timescale of some tens of Myrs when 
the perturber is a Jovian--type planet limiting the applicability 
of our computations to older disks. 
From Eq. (\ref{A2}) we can deduce that 
smaller size planets need more time to randomise the pericentre 
since the secular period is inversely proportional to the planet mass
$m_{pl}$. For example, a Neptune--size planet would lead to 
pericentre randomisation on a timescale approximately 20 times longer
i.e. about half of a Gyr. Our estimates of 
$P_i$ and
$U_c$ will be valid only after that time.

\section{The method}

An essential prerequisite to model the collisional evolution of 
a population of minor bodies,
like planetesimals, is a quantitative estimate of two parameters: 
the average impact velocity $U_c$ and the intrinsic probability of 
collision $P_i$. The impact velocity appears in all basic steps 
determining the outcome of a collision and it discriminates 
between accumulation, cratering or shattering events. Together with 
semi-empirical scaling laws describing the strength of a body suffering a 
collision,  $U_c$  determines the mass of the largest remnant 
body and the size distribution of the escaping fragments. 
The intrinsic probability of collision $P_i$ is related to   
the frequency of collisions between the bodies of the given population.
It is a property related to the orbital distribution and defined 
independently from  
the number or size distribution of the bodies populating the belt. 
For example, for two given orbits, $P_i$ is defined as the mean number per year of close
encounters with minimum distance less than $1$ km occurring between two points
moving along the two orbits. If two bodies with radius $R_T$ (``target'') and 
$R_p$ (''projectile'') move along the above mentioned orbits,
the mean number of collisions per unit of time is $P_i(R_T+R_p)^2$.
In wider terms, if the orbit of a target body with radius $R_T$ moves within a swarm of $N_p$ projectiles
with radius $R_p$, and $P_{i,k}$ is the intrinsic probability of collision between the
target orbit and the orbit of the $k$-th projectile, the mean number of collisions per unit of time
suffered by the target is $\sum_k P_{i,k}(R_T+R_p)^2$, or:

\begin{equation}
\frac{{\rm d} n}{{\rm d} t} = \langle P_i \rangle (R_T + R_p)^2 N_p
\label{Pi}
\end{equation}

\noindent
where $\langle P_i \rangle = (\sum_k P_{i,k})/N_p$ is the {\it mean} intrinsic probability of collision
\citep{fari92,davis02}.
In this paper we will treat only cases of targets impacted by populations of projectiles, 
so hereinafter we use the symbol $P_i$ always with the meaning of mean intrinsic probability of collision.
It has been computed for 
different populations of  minor bodies in the solar system
and, for instance, its mean value for the asteroid belts is 
$P_i \sim 2.9 \times 10^{-18}$ km$^{-2}$ yr$^{-1}$ 
\citep{Bottke16}. The important aspect of $P_i$ is that it
is not simply a particle--in--a--box model where the 
bodies move freely between collisions, but it accounts for the 
Keplerian dynamics of the bodies even when they are on 
perturbed orbits. 

In the case of planetesimal belts perturbed by a planet in a close-by orbit, 
the classical \"Opik/Wetherill  approach for the computation
of $P_i$ and $U_c$ cannot be used if the forced eccentricity is not
significantly smaller than the proper eccentricity. 
The \"Opik/Wetherill statistics, hereinafter 
termed  ``canonical'' statistics, is based on the following assumptions:

\begin{itemize}
\item[(1)] the semimajor axes $a$, eccentricities $e$ and inclinations $I$ of the osculating orbits are fixed;
\item[(2)] the rate of variation of the node longitudes $\Omega$ and pericentre arguments
$\omega$ of the osculating orbits are constant. In other words, the osculating
orbits circulate uniformly;
\item[(3)] the motion of the planetesimals along their osculating orbits is fully described by 
the second Kepler's law;

\end{itemize}

The first attempt to overcome the assumptions of the canonical statistics has been done by
\citet{delloro_paolicchi_1997, delloro_paolicchi_1998,delloro_paolicchi_1998b}. 
In particular \citet{delloro_paolicchi_1998}
developed a mathematical formalism for the study of the statistics of collisions among asteroids
provided that the following more general assumptions are fulfilled:

\begin{itemize}
\item[(1)] the orbital parameters $a$, $e$ and $I$ of the osculating orbits are fixed
\item[(4)] none of the osculating elements $a$, $e$ and $I$ is correlated with one of
the angular elements $\Omega$, $\omega$ or mean anomaly $M$
\end{itemize}

Indeed, thanks to its numerical implementation, the method of \citet{delloro_paolicchi_1998} can be used even if condition (1) 
is not fulfilled by simply substituting 
the ensemble of fixed values of $a$, $e$ and $I$ with a fictitious list of ``child'' elements $a$, $e$ and $I$ 
whose distributions  fit {\it ad-hoc} parent distributions.
So, in general, in all cases where condition (4) is fulfilled, 
the method proposed by \citet{delloro_paolicchi_1998}
can be employed.
Unfortunately, the dynamical behaviour of a planetesimal belt perturbed by a planet  
strongly violates condition (4) because of the strong correlation between eccentricity and 
perihelion longitude, entailing a correlation among $e$, $\Omega$ and $\omega$.
To account for this correlation, we envisioned a different technique described here below. 

Our new algorithm adopts a Monte Carlo approach consisting in a random exploration 
of the phase space of positions and velocities of the bodies in order to derive
their rate of collision and impact velocity distribution. It
can be summarised in four consecutive steps. 

In the first step, a numerical model of the disk is prepared by generating a 
list of bodies with semimajor axis randomly and uniformly chosen in 
an interval $[a_{min}, a_{max}]$. 
Different radial distributions can be implemented, but in this paper we adopt 
this simple assumption which implies a decrease of the planetesimal 
superficial density as $r^{-1}$. 
The forced eccentricity for each body
is computed from Eq. \ref{ef} \citep{mustill2009} while 
the inclination $I$ is randomly generated from a uniform distribution 
in the interval $[0,I_{max}]$.
Finally, a value of proper eccentricity $e_p$ is computed for all bodies which is derived 
from the initial assumed value of proper eccentricity $e_{p0}$ of each planetesimal 
in the belt before 
it is perturbed.
In all our models, the average number of bodies (orbits) used in the Monte Carlo statistics
 is of the order of
$10,000$. Our final goal is to compute the statistical parameters $P_i$ and $U_c$ for collisions
occurring between 
all bodies belonging to our list, considered representative of the structure of the disk, and some selected 
target bodies (tracers) whose orbits have inclination $I_T=I_{max}/2$, 
semimajor axes $a_T$  chosen at fixed steps between $[a_{min}, a_{max}]$, 
and values  of $e_p$, $e_f$
computed 
with the same rules for the disk bodies. 
In this way, $P_i$ and $U_c$ are functions of the semimajor axis of the target only
and can be evaluated at different locations on the disk. In the case of 
a warm belt, since the tracers may have different values of $e_p$, 1000 tracers
are used for each semimajor axis and the values of $P_i$ and $U_c$ are computed 
as average over all of them. 

The second step consists of a random sampling of the mean anomaly $M$
and of the secular  
angle $\theta$, the angle between the vectors describing the proper and forced eccentricity,
respectively
(see Fig. 1). Both samplings are performed assuming a uniform distribution of 
the  angles since we know from the secular theory that  $\theta$ precesses at
a regular pace, at least as a first approximation,  while the mean anomaly $M$ 
circulates with the constant frequency 
$n$. From the values of $e_f$, $e_p$ and $\theta$ we can derive 
the osculating eccentricity completing the set of orbital elements. 
From the long list of orbital elements sampled for each body of the disk  and 
tracers we can 
compute the list of positions 
and velocities $({\bf r}_t, {\bf v}_t, {\bf r}_p, {\bf v}_p)$  of projectiles and targets. 
This two-steps sampling allows to reproduce in the phase space, 
described by positions and velocities, the distribution of the orbital 
elements of projectiles and targets imposed by the secular perturbations of the planet.
This distribution is characterised by different levels of   
correlation between $e$ and $\tilde\omega$ which depend on the initial choice of 
$e_{p0}$.  {\it At this point we have a model 
disk reflecting the secular dynamics in Cartesian coordinates.} 

The third step consists in computing, from the distribution of the positions and velocities of both projectiles and targets, the rate of close approaches and the 
distribution of impact velocities. 
For each target the number $\nu(R)$ of close encounters within a given distance $R$ per unit time is, by definition,
the ratio:

\begin{equation}
\nu(R) = \frac{N(R,T)}{T}
\end{equation}
where $N(R,T)$ is the number of close encounters occurred during an interval of time $T$ and within a distance $R$.
It is noteworthy that, in the context of the above equation, $R$ 
is the distance between the centres of the two bodies during the approach and it is not the 
radius of either of the two bodies.
Assuming a general dynamical stability of the system and choosing $T$ long enough, the number $\nu(R)$ 
no longer depends either on the initial conditions of the system or on the value of $T$. We can rewrite this
ratio as:

\begin{equation}
\nu(R) = \frac{N(R,T)}{S(R,T)}\frac{S(R,T)}{T} 
\end{equation}
where $S(R,T)$ is  the sum of the durations of all close encounters occurred during the interval 
of time $T$ and within the distance $R$. The ratio:

\begin{equation}
\tau(R) = \frac{S(R,T)}{N(R,T)} 
\end{equation}
is the mean value of the durations of the close encounters within a distance $R$, while:

\begin{equation}
p(R) = \frac{S(R,T)}{T} 
\end{equation}
is the probability to find a projectile within a distance $R$ from the target at a randomly chosen instant of time.
The mean close encounter duration $\tau(R)$ is related 
to the properties of the relative motion between projectile and target. Assuming that the 
relative trajectory of the bodies  can be approximated as a rectilinear motion with constant velocity $v = |{\bf v}_t - {\bf v}_p|$, the average (expected) duration of a
close encounter is $(4/3)(R/v)$, taking into account that
the projectile can pass anywhere inside the sphere of radius $R$ around the target.
This rectilinear motion approximation is valid only for values of the close approach distance $R$
significantly smaller than the size of the orbits of the planetesimals.
The probability $p(R)$ is directly derived  from the $({\bf r}_t, {\bf v}_t, {\bf r}_p, {\bf v}_p)$ list
by computing the ratio between the number of samples for which $|{\bf r}_t - {\bf r}_p| < R$ and the total number of
samples.
In this way the rate of the close encounters can be expressed in terms of the sampled quantities as:

\begin{equation}
 \nu(R) = \frac{3}{4}\frac{1}{{\cal N}R}\sum_k v_k 
\label{nuR}
\end{equation}
where $v_k$ is the relative velocity of the $k$-th sample of target-projectile pairs.
The sum includes only those cases for which $|{\bf r}_t - {\bf r}_p| < R$ while ${\cal N}$ is the total
number of samples \citep{oro2016}.

The rate $\nu(R)$ is derived as a 
function of the close approach distance $R$.  
By definition of intrinsic probability of collision,
$P_i = \nu(R)$ if $R=1$ km. 
On the other hand, due to the numerical limitations of the Monte Carlo approach, 
it is not possible to evaluate the rate $\nu(R)$ directly for $R=1$ km, since  it requires an excessive 
computational effort. 
Nevertheless, for values of $R$ small compared to the 
linear dimensions of the orbits, the rate
$\nu$ is expected to be proportional to the geometric cross section, that is $\nu(R) \propto R^2$ (the effect of the gravitational 
focusing is negligible due to the small sizes of the planetesimals). 
For all the cases investigated in this work we have verified that $\nu(R)$ is really proportional 
to $R^2$ when $R$ is small enough. For this reason, we extrapolate 
the function $\nu(R)$ down to $R=1$ km by assuming that the $R^2$ trend is maintained.
In short, within the interval of values of $R$ for which $\nu(R)$ results to be proportional to $R^2$, the ratio 
$\nu(R)/R^2$ provides the value of $P_i$ directly.

Together with $P_i$, we also compute 
the mean $\overline{v}(R)$ of the relative velocity for all close encounters with $|{\bf r}_t - {\bf r}_p| < R$.
The parameter $\overline{v}(R)$ is correctly evaluated on the basis of the $({\bf r}_t, {\bf v}_t, {\bf r}_p, {\bf v}_p)$ list
as:
\begin{equation}
\overline{v}(R) = \frac{\sum_k v_k^2}{\sum_k v_k} 
\label{vmR}
\end{equation}
where again $v_k$ is the relative velocity of the $k$-th target-projectile pair and the sum includes only those cases for which $|{\bf r}_t - {\bf r}_p| < R$.
From Eq. \ref{nuR} it can be deduced that 
each of the target-projectile pairs contributes to the final evaluation of the frequency $\nu(R)$ with 
a term $v_k/R$. This comes from the fact that each orbital geometry has its own probability to occur, as outlined 
by \cite{bottke94}. In other words, 
each value of the relative velocity $v_k$ of the planetesimal ensemble has to be weighted by $v_k$ itself 
in order to obtain the correct probability distribution of the relative velocities.
The same holds true for any other parameter related to close
encounters.
But unlike $\nu(R)$, the value of $\overline{v}(R)$ is expected to tend asymptotically to a finite value for $R$ 
smaller and smaller. This limit is the mean value $U_c$ of the impact velocity. In general, for the 
cases under investigation in this work, we have verified that the function $\overline{v}(R)$ decreases for
lower $R$ and it becomes constant from a certain value onward, providing our estimation
of $U_c$.

The procedure described above
leads to an accurate estimate of both $P_i$ and $U_c$ \citep{oro2016}.
The algorithm has been extensively validated on different
sets of test
cases and, in particular, it provides values of $P_i$ and $U_c$ 
for Main Belt asteroids and Kuiper belt objects which are in agreement with those
previously reported in literature.

The direct source of uncertainty in the computation of $P_i$ and $U_c$ is
due to the casual fluctuations in the random sampling of the phase 
space, depending on the total number of samples (and so the total duration 
of the computation). Another, but indirect, error is introduced by the 
fluctuations in the construction of the ring model. The list of planetesimal orbits 
is generated randomly and not all the particles in the ring intersect
the orbit of the target, but only a fraction of them within a more or less wide range of 
semimajor axes. This means that, depending of the details of the random generation
of the orbits, the number of bodies intersecting the orbit of the target
can change a little, impacting on the evaluation of $P_i$ and $U_c$. The 
total number of random samples has been tuned in order to constrain the uncertainties
within few percents of $P_i$ and $U_c$.

\section{The "standard" case}

As a first test, we apply our algorithm to compute the intrinsic 
probability of collision and average impact velocity to a "standard" case
similar to that explored by \cite{mustill2009}. In this model, a planetesimal
belt extends from 10 to 20 AU and it is perturbed by a Jupiter--sized
planet with mass $m_p = 0.001$ $m_{\odot}$ on an orbit with semimajor axis $a_{pl} = 5$ AU
while the eccentricity can be either $e_{pl}= 0.1$  or $e_{pl} = 0.6$. 
We consider two distinct configurations: a 
cold belt that, before the arrival of the planet in a perturbing orbit, had 
a very low proper eccentricity ($e_{p0} \sim 0$). 
In this case, when the planet turns on its secular perturbations, the value of 
$e_p$ becomes approximately equal to $e_f$ (see Fig.\ref{secular}). We also explore the evolution of 
warm belts that, prior to the planet arrival, had already a high value
of proper eccentricity $e_{p0}$, possibly excited by other mechanisms like an 
extended period of planet--planet scattering or the formation of large 
embryos.  In this scenario we assume that $e_{p0}$ is  a significant 
fraction of $e_f$ and we test three cases, one with $e_{p0} = 0.25 \cdot e_f$,
one with $e_{p0} = 0.5 \cdot e_f$ and one, the most excited, with $e_{p0} = e_f$.
This is just an arbitrary choice to 
show the effects of an initial value of $e_{p0} > 0$ and it is not dictated 
by any particular scenario. It appears a better choice than a random selection of 
$e_{p0}$ values. As soon as the planet begins to perturb the 
warm belt, the eccentricity values will be encompassed between 
$e_f - e_p$ and $e_f + e_p$ where $e_p$ is derived from the initial 
value of $e_{p0}$ as described in Sect. 2 (see Fig.\ref{secular2}). 

\subsection{Initially cold belt}

\begin{figure}
  \includegraphics[width=\hsize]{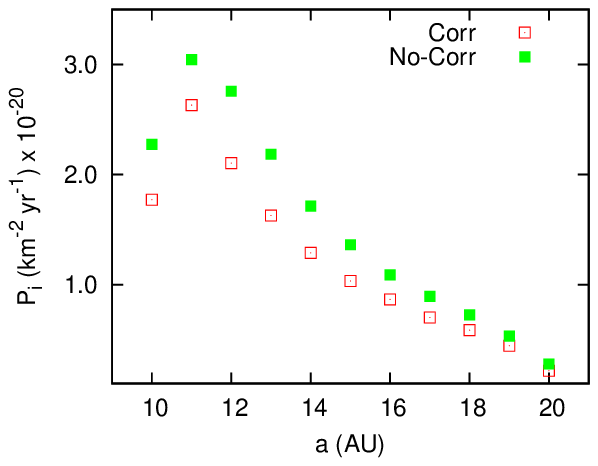}
  \includegraphics[width=\hsize]{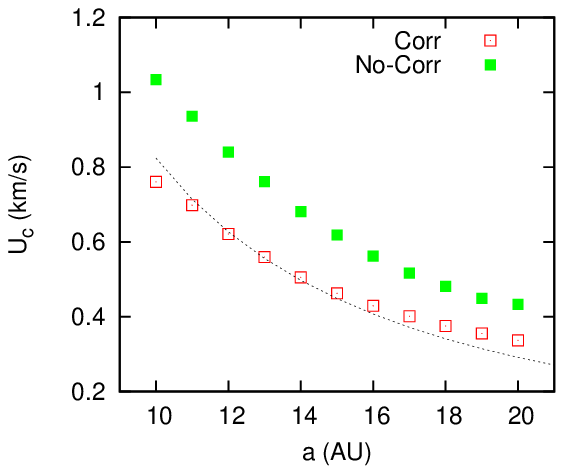}
\caption{\label{apl5e01} Intrinsic probability of 
collision $P_i$ (top panel) and impact velocity $U_c$ 
(bottom panel) for a planetesimal belt perturbed 
by a Jupiter--size planet with $a_p = 5$ and $e_p= 0.1$. 
The belt is assumed to have been initially cold so that 
$e_p = e_f $. The green filled squares mark the 
predictions of the canonical model where no secular
correlation is assumed between $e$ and $\varpi$ while 
the empty red squares are the values computed with the model 
that include the secular correlation.
The dashed line in the bottom 
panel outlines  the values derived from 
the analytic formula $U_c = 1.4 e_f v_{kep}$ \citep{mustill2009}.
}
\end{figure}

In an initial cold belt the correlation between 
$e$ and $\varpi$ that the planet establishes when approaching the 
belt is the strongest due to the assumption that 
$e_{p0} = 0$. The secular dynamics is well described by the 
plots of Fig.\ref{secular}. Both $P_i$ and $U_c$ 
are shown in Fig. \ref{apl5e01}  
as a 
function of the 
semimajor axis of the tracers which well approximate the average
radial distance from the star of each body. 
In this figure we illustrate the case with 
$e_{pl}= 0.1$, a configuration in which the planetesimal belt is 
less perturbed (compared to $e_{pl}= 0.6$). 
The green full squares are 
the impact velocity values computed assuming there is no secular correlation 
between the eccentricity and perihelion longitude (hereinafter the 'canonical' case)
while the empty red squares 
show the probability and velocity values when the secular correlation is taken into account. 
The dotted line in
the lower panel of Fig. \ref{apl5e01} gives  
the average relative velocity computed as $U_c = 1.4 e_f v_{kep}$ 
where $v_{kep}$ is the local Keplerian velocity \citep{mustill2009}.  
This relation leads to a simple $r^{-3/2}$ scaling of the 
velocity with the radial distance $r$.
The correlation between eccentricity and perihelion 
longitude reduces both $P_i$ and, in particular, the relative 
impact velocity $U_c$ 
respect to the case without secular correlation. 
This is due to two effects which are manifest in Fig. \ref{secular}: there is a consistent 
fraction of orbits with very low eccentricity while the high eccentric orbits 
have their pericentres almost aligned around $0^o$ leading to low velocity impacts.

The analytic 
prediction of \cite{mustill2009} for $U_c$ appears to slightly overestimate the impact 
speed at 10 AU by about 10\% and underestimate it at 20 AU
by approximately the same amount.   The $r^{-3/2}$ scaling does not 
appear to fully account for the dynamical evolution of the belt 
possibly because it does not properly account for the spatial 
distribution of the encounters radial locations related to the secular dynamics. 
For $P_i$, it is difficult to derive a proper scaling with the 
radial distance since two different effects come into play in 
determining $P_i$ as a function of $r$. On one side there is a pure 
dynamical dependence of the probability of collision of a 
planetesimal pair on $r$ related to their orbital elements.
A typical equation giving such a probability can be found in \cite{wetherill67} 
(eq. 20). However, we are computing the $P_i$ not of a single pair, 
but for an entire population of planetesimals and their radial distribution 
comes strongly into play. This radial distribution 
does not only depend on the planetesimal superficial density 
distribution but it is significantly 
influenced by the distribution of all orbital elements of the population.
As a consequence,  it cannot be 
predicted a priori with an easy power law distribution, in particular
for highly eccentric belts.

Close to the inner edge of 
the belt, the value of $P_i$ decreases 
due to the inner truncation of the belt and to a reduction in
the number of crossing orbits. This effect will disappear 
for higher eccentricities since the perihelion--aphelion radial 
distance is larger for the planetesimals and the orbital crossing is more 
extended. 

\begin{figure}
  \includegraphics[width=\hsize]{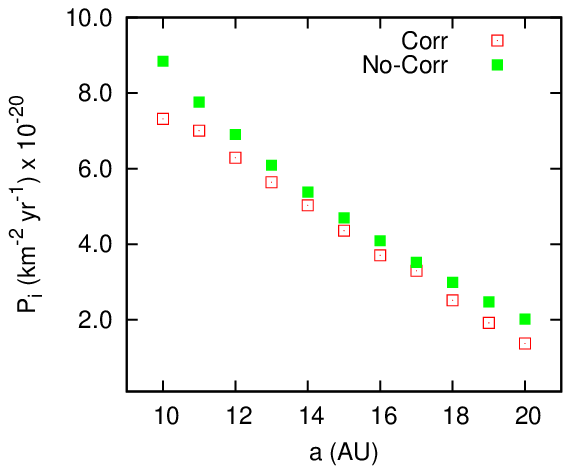}
  \includegraphics[width=\hsize]{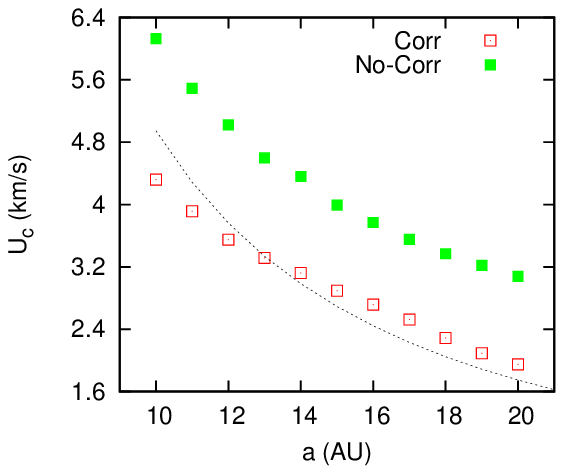}
\caption{\label{apl5e06} Same as Fig. \ref{apl5e01}
but 
for $e_p= 0.6$. 
}
\end{figure}

In Fig. \ref{apl5e06} we show both $P_i$ and $U_c$ when 
the eccentricity of the planet is increased to 0.6, 
a significantly more perturbed configuration for the 
planetesimal belt. 
The value of $P_i$ is increased by about a factor 4 leading to a 
very active belt in terms of collisions while the impact velocity 
$U_c$ is five times higher compared to the case with 
$e_{pl} = 0.1$  and comparable to  the value of  the present  
asteroid belt. The vast majority of collisions are expected to 
be highly energetic 
and both fragmentation and cratering  are possibly dominant leading to 
a high rate of dust
production.  These results show that a higher eccentricity of the planet
not only leads to higher impact speeds but it substantially increases 
the impact rate (larger $P_i$) which is possibly more important in producing 
brighter debris disks.

\begin{figure}
  \includegraphics[width=\hsize]{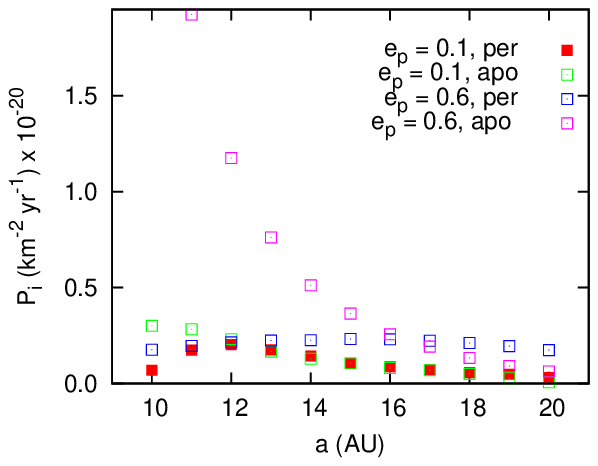}
  \includegraphics[width=\hsize]{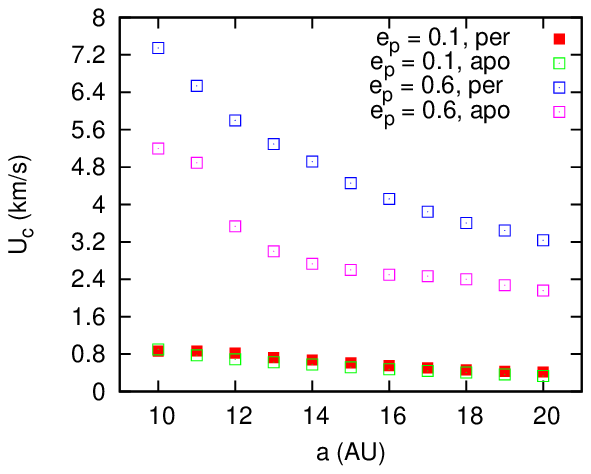}
\caption{\label{peri} Intrinsic probability of
collision $P_i$ (top panel) and impact velocity $U_c$
(bottom panel) computed around the pericentre  
($-15^o < f < 15^o$) and apocentre ($165^o < f < 195^o$) 
of all targets. The eccentricity of the planet is set to 
$e_p = 0.1$ and $e_p=0.6$. 
}
\end{figure}

Our numerical algorithm allows to compute the contribution 
to the
average $P_i$ and $U_c$  coming from  a  restricted arc of the trajectory of  the
target, or,  more precisely,  from a  given range of  values of  the true
anomaly $f$. 
Thanks to this feature, we can evaluate  the frequency of collision and 
impact velocity around the pericentre and apocentre, respectively.
We have selected a range of $\pm 15^o$ around both 
$f = 0^o$ and $f = 180^o$ for each target in order to underline different 
values of $P_i$ and $U_c$ in these restricted ranges. The results are shown in Fig. \ref{peri} 
for the models in which the eccentricity of the planet is set to 
$e_p = 0.1$ and $e_p=0.6$, respectively. 

When $e_p = 0.1$, the difference between $P_i$ and $U_c$ at pericentre 
and apocentre is small. However, when $e_p = 0.6$ a significantly
higher value of $P_i$ is observed when the target is at the apocentre and hence
in the inner regions of the belt. Beyond 17 AU  the trend reverses and  
$P_i$  becomes larger at pericentre. 
 This behaviour is due 
to the high forced (and then proper) eccentricity induced by the secular
perturbations of the planet. When the target is close to the inner border of 
the belt its pericentre is located where the density of 
potential projectiles is significantly lower due to the overall eccentricity 
distribution ranging from 0 and $2 \ e_f$.  On the other hand, when the target orbits 
around the apocentre, it is well within the belt where the 
density of the projectiles is the highest and $P_i$  is large.
Moving towards the outer border of the belt, it is now the apocentre of the targets 
that is located  in low density regions while the pericentre is well within the belt 
and this leads to higher values of $P_i$. 
Superimposed to this effect there is 
also the decreasing trend of $P_i$ for larger values of $a$ due to a reduction of 
both the forced eccentricity and radial density of the planetesimal population. 
The values of $P_i$ shown in  Fig. \ref{peri} are only a fraction of the 
total impact probability illustrated in Fig. \ref{apl5e06} since we 
are considering only a portion of the total range of the true anomaly $f$ 
and, as a consequence, they are substantially smaller than the values in Fig. \ref{apl5e06}.

Significantly different values of $U_c$ are also found at pericentre and 
apocentre when $e_p = 0.6$.
At apocentre the values of the impact speed 
are comparable with the average values shown in Fig. \ref{apl5e06}
but at pericetre $U_c$ is much higher since the 
target has a higher orbital velocity compared to the potential projectiles.
Due to the correlation between $e$ and $\varpi$, a significant number of impacts 
when the target is at the pericentre occurs with circular orbits having radius 
approximately equal to $a (1 -e)$. By comparison, at apocentre the target will 
frequently encounter projectiles on circular orbits with radius $ a(1 + e)$. 
As a consequence, at pericentre the impact velocity $U_c$ is higher. 

\subsection{Initially warm belt}

In a scenario where the planet is injected in an orbit perturbing an 
already heated planetesimal belt, the values of $P_i$ and $U_c$ are different
respect to those of a cold belt because of 
the changes in the secular dynamics (Fig. \ref{secular2}).
Here we consider three test cases where 
the proper eccentricity $e_{p0}$,  before the onset of the 
planet perturbations, is 0.25, 0.5 and 1.0 that of the forced eccentricity
(the value that will be established once the planet approaches the belt). 
Different values of $e_{p0}$ can be encountered in real systems and our choice to 
link $e_p$ to $e_f$ is arbitrary, but it is impossible to explore all 
possible initial distributions of $e_{p0}$, so we consider only three reference cases,
where $e_{p0}$ is linearly related to $e_f$, which give 
clues on how to deal with any general scenario. 

\begin{figure}
  \includegraphics[width=\hsize]{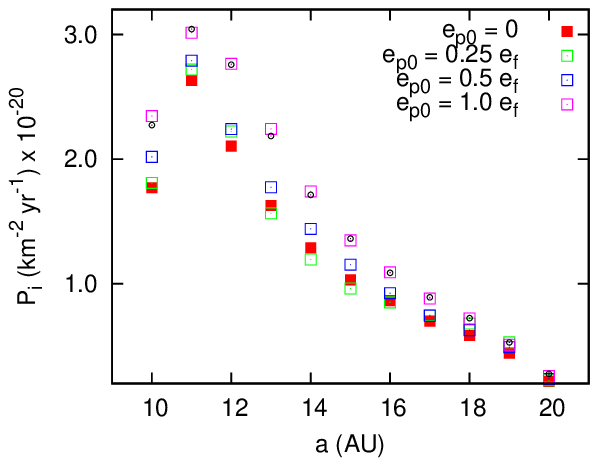}
  \includegraphics[width=\hsize]{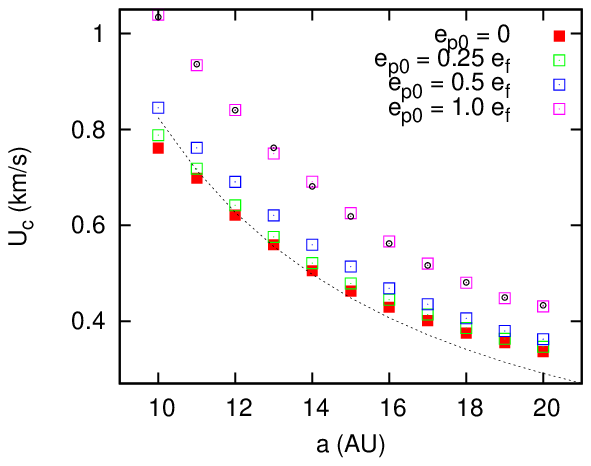}
\caption{\label{apl5e01epef} $P_i$ (upper panel) and $U_c$ (lower panel) for initially 
warm and cold  belts. 
For the warm belts we assume three 
different values of $e_{p0}$, i.e. 0.25 (green empty squares), 0.5 (blue empty squares)  and 1 times $e_f$
(magenta empty squares). The red filled squares mark the values of  $P_i$ and 
$U_c$ for a cold belt while the black empty circles show the case with $e_p = e_f$ {\it without} 
secular perturbations. 
The eccentricity of the perturbing planet is set to $e_{pl} = 0.1$.
}
\end{figure}

\begin{figure}
  \includegraphics[width=\hsize]{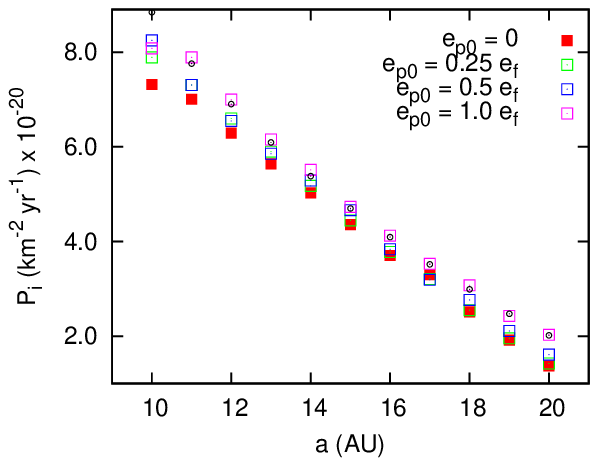}
  \includegraphics[width=\hsize]{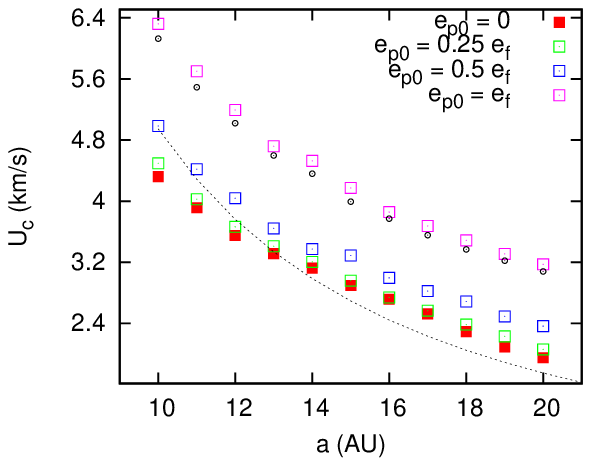}
\caption{\label{apl5e06epef} Same as Fig. \ref{apl5e01epef}  
but for $e_{pl} = 0.6$.
}
\end{figure}

In Fig. \ref{apl5e01epef} 
we compare $P_i$ and $U_c$ when $e_{pl} = 0.1$ for all the  different initial
configurations corresponding to distinct dynamical states of the belt prior to the 
onset of the planet perturbations,  i.e. cold and warm belts. This comparison highlights 
the effects on $P_i$ and $U_c$ of different degrees of correlation  
between $e$ and $\varpi$ related to the distinct initial values of $e_{p0}$. 
Both the intrinsic 
probability of collision and impact velocity increase for higher values of $e_{p0}$ 
(initially warm belts) until, for $e_{p0} = e_f$, values comparable to those of 
the uncorrelated case (the canonical one) are obtained. This increase is due to the complex interplay of  
dynamical effects which can be deduced from Fig.\ref{secular2}.   
First of all in warm belts higher eccentricity values 
are achieved
since $e_p$ ranges from
$e_f - e_{p0}$ to $e_f + e_{p0}$  respect to the cold scenario where $e_p = e_f$.
In addition, when the
pericentre $\varpi$ is around $180^o$ the planetesimal eccentricities
are not all small as
illustrated in Fig.\ref{secular} but they may reach significant values
(Fig.\ref{secular2}).
Finally, pseudo--librators, whose fraction depends on the initial value of 
$e_{p0}$, 
impact with non--librators leading to geometrical configurations where the
trajectories at the crossing  are more bent and favour larger collisional velocities.
All these mechanisms contribute to increase both $P_i$ and $U_c$ up to 
values similar to those of the canonical case also 
when $e_{pl} = 0.6$, as illustrated in Fig. \ref{apl5e06epef}. 

Our numerical models show that initial warm belts have a more intense collisional activity due to higher 
values of both $P_i$ and $U_c$.  This leads to a greater production of dust and 
a brighter debris disk associated to the belt but also to a faster erosion rate with a
shorter lifetime.

\begin{figure}
  \includegraphics[width=\hsize]{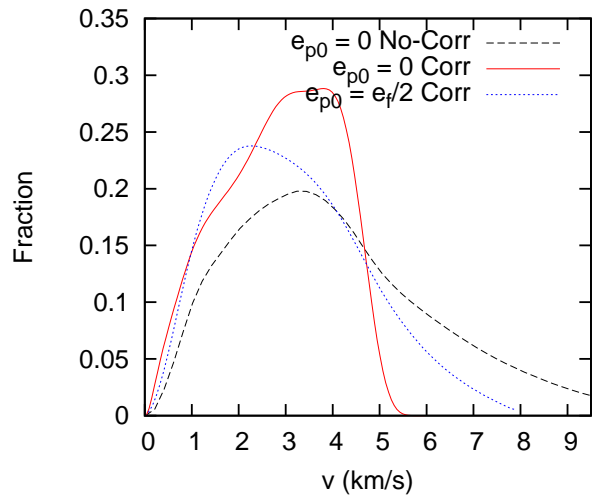}
\caption{\label{distri} Normalised impact velocity distribution for $e_p = e_f$
without the secular correlation between $e$ and $\varpi$ (black dashed line) and for
the case in which such correlation is included in the computations (red
continuous line). The blue dotted line illustrates the velocity distribution 
for a warm belt with 
$e_{p0} = 0.5 \cdot e_f$.  
}
\end{figure}

The value of impact velocity $U_c$ shown in the previous plots is indeed an
average over a large number of computed impact speeds. The velocity 
distribution in the standard case with $e_{pl}= 0.6$ for the target with $a_T=15$ AU is displayed in  
Fig. \ref{distri}. The black dashed line shows the impact velocity 
distribution in absence of $e$--$\varpi$ correlation (a warm belt 
not perturbed by a planet) 
while the red continuous line illustrates the distribution when such 
correlation is included in the computation of $U_c$ for an initially cold belt. The high velocity tail of the 
distribution is cut off and this explains the reduction in the average
impact speed observed in Fig.\ref{apl5e06}. In 
the warm case  with $e_{p0} = 0.5 e_f$  the cut off at high impact speeds is reduced 
compared to the case with $e_{p0} = 0$ 
due to the contribution of pseudo--librators impacting circulators. 
However,  the peak is located at lower impact speeds and the 
average value of $U_c$ is still lower compared to the case without 
$e$--$\varpi$ correlation.

\subsection{Initially warm and inclined belt: $i = 15^o$}

\begin{figure}
  \includegraphics[width=\hsize]{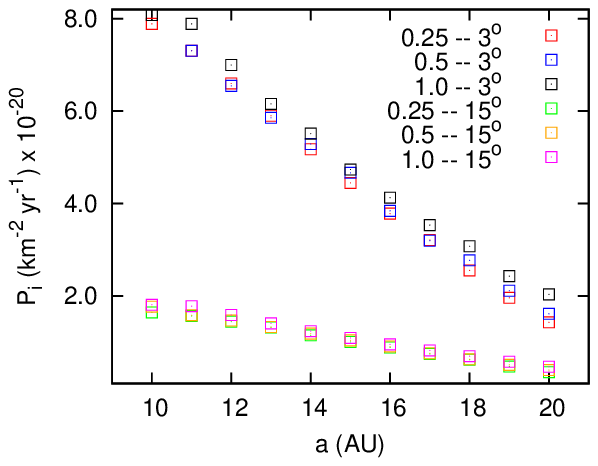}
  \includegraphics[width=\hsize]{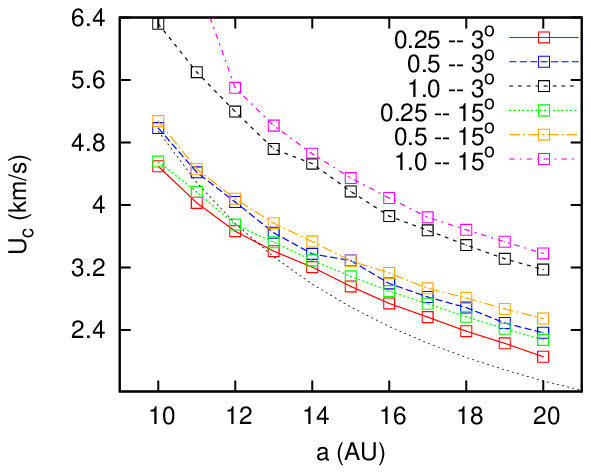}
\caption{\label{incli1}  Intrinsic probability of 
collision $P_i$ and impact velocity $U_c$ for a small body 
belt with $e_{p0} =$ $0.25$, $0.5$ and $1.0$  $e_f$ and $i = 15^o$,  compared 
to the values computed for $i=3^o$. The perturbing 
planet has an eccentricity $e_{pl} = 0.6$. While $P_i$ is 
significantly reduced, the impact velocity $U_c$ is 
approximately increased by 40\%. 
}
\end{figure}

In presence of a strong dynamical excitation, due to possible different 
mechanisms like self--stirring, the presence of large planetary embryos, 
resonance sweeping etc., even the 
osculating inclinations may be significantly pumped up.
We explore in this section the effects of increasing the 
average value of inclination on the 
values of $P_i$ and $U_c$. Intuitively, one would expect an 
increase in the relative impact velocity due to the presence of an additional 
out--of--plane component in the difference between the velocity vectors 
of two planetesimals. At the same time, due to an expansion of the 
available space for orbital motion, a significant decrease 
of the impact probability is expected. Both these predictions are 
confirmed by the outcome of our algorithm and in 
Fig. \ref{incli1} we show the computed values of $P_i$ and $U_c$ 
for initially warm belts with $e_{p0} = 0.25, 0.5,  1  \cdot e_f$ and an average 
inclination of $15^o$ (the perturbing planet has $e_{pl} = 0.6$). 
The exploration of larger values of inclination would require a more 
detailed secular approach where the inclination is coupled to the 
node longitude and this will be done in a forthcoming paper.

In the case of inclined belts, the value of $P_i$ in the inner regions of the 
belt is reduced by about a 
factor of 4  compared to the case with $i=3^o$ 
(see Fig.\ref{incli1}). 
On the other hand, the impact velocity $U_c$ is higher for 
more inclined planetesimals due to the out--of--plane component in the 
relative velocity, reaching a maximum value of 8 km/s at 10 AU from the 
star when $e_{p0} = 1 \cdot e_f$ (magenta empty squares in the bottom panel of  Fig. \ref{incli1}). 
However,  a higher inclination appears to be not as important as 
a higher values of $e_{p0}$ in leading to higher impact speed. 
In general, according to our modelling, 
an inclined belt is expected to be 
less collisionally active due to the strong decrease in $P_i$, in spite of 
an increase of $U_c$. This 
will be confirmed by the collisional evolution models of the next section.

\subsection{Collisionally damped  belt}

We consider here the case of a densely populated belt whose 
proper eccentricity distribution, after the approach of the planet,
has been collisionally damped. In this scenario most of the 
orbital excitation is dissipated and
the proper eccentricity is reduced to values smaller than the forced one.
This would lead to the dynamical state illustrated in 
Fig.\ref{secular3} where all the planetesimals are in 
pseudo--libration.
We compute the intrinsic probability of collision and impact velocity when
the damped configuration is reached. To model this kind of belt 
we select a value of proper eccentricity, after the damping, which is half the
value of the forced one. Even in this case, the choice is made to
give an idea of the influence of this secular configuration on the
collisional evolution of the belt, i.e. on the values of $P_i$ and $U_c$,
and it is not related to any particular scenario.

\begin{figure}
  \includegraphics[width=\hsize]{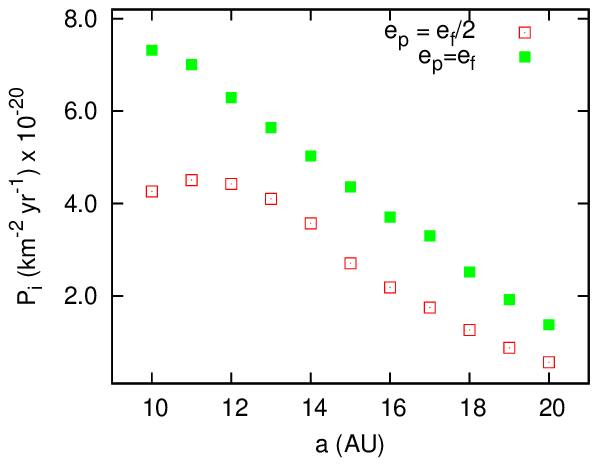}
  \includegraphics[width=\hsize]{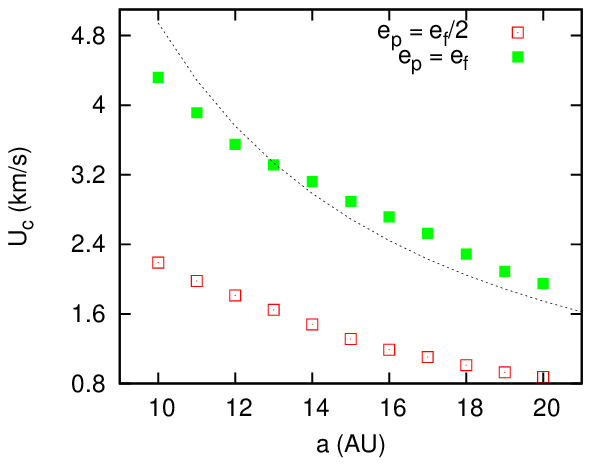}
\caption{\label{damp}  Intrinsic probability of
collision $P_i$ and impact velocity $U_c$ for a small body
belt collisionally damped after the onset of the 
planet perturbations. The eccentricity of the planet is 
set to 0.6 and the proper eccentricity is assumed to have 
been damped to a value $e_p = 0.5 \cdot e_f$.
Both $P_i$ and $U_c$ are compared to the values computed 
for a non--damped cold belt (filled green squares). 
There is a significant decrease of both $P_i$ and $U_c$ due 
to the pseudo--libration of all planetesimals. 
}
\end{figure}

For a damped belt both $P_i$ and $U_c$ are 
reduced (see Fig.\ref{damp}) and this has to be
ascribed to the pseudo--libration of all planetesimals. $\varpi$ oscillates within 
a limited range around $0^o$ leading to a significant level of pericentre 
alignment of all orbits. This dynamical configuration leads to collisions where 
the trajectories of the approaching planetesimals are almost parallel 
at the orbital crossing. As a consequence, there  is a consistent reduction of 
the impact velocity and impact rate. A damped belt is then 
expected to be less collisionally active and it will give rise  to debris disks which 
are less bright. 

\section{Collisional evolution}

To explore the implications of different values of $P_i$ and $U_c$ on the 
evolution of a planetesimal belt we have run a 
simple
one-dimensional collisional evolution code \citep{campo94, marza95, bottke2005}.
We start from an initial planetesimal population extending from 
10 to 500 km in diameter and distributed in a series of discrete 
logarithmic diameter bins following a slope equal to -3.5. The
population is centred at 15 AU and it is 10 AU wide.
The code computes the collisional interactions of each
size bin with every other one during a given time--step. At the end 
of the time--step, all the outcomes of the interactions are summed up to find the 
net change in the population as a function of size. The updated population 
is then used in the next time--step until the whole time--span, that we 
assume to be 4 Gyr, is covered. These calculations are performed under the 
assumption that the orbital element distribution of the planetesimals 
is not significantly altered by the collisions so that both 
$P_i$ and $U_c$ remain constant during the entire evolution of the belt. 
Two different scaling laws for the computation of the fragments size 
distribution after a catastrophic impact are tested: a simple 
energy scaling like that proposed in \cite{faridav97} to study the collisional 
evolution of Kuiper Belt objects  and the more recent one 
given in \cite{stew2009} for weak aggregates (their Eq. 2). 
The results are not significantly different so we report only those 
obtained with the scaling law of \cite{stew2009}. 

\begin{figure}
  \includegraphics[width=\hsize]{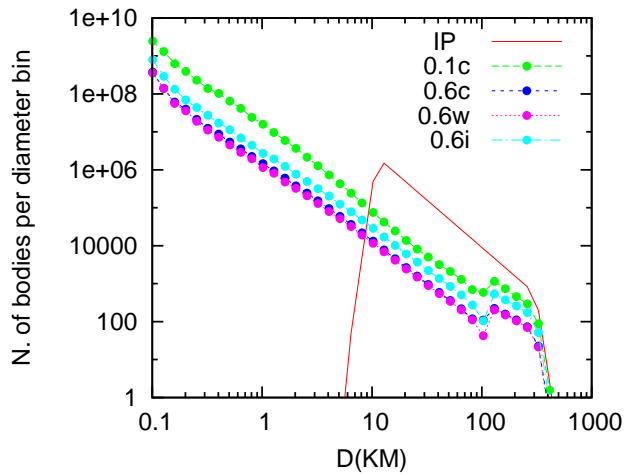}
\caption{\label{collev} Collisional evolution over 4 Gyr of a putative planetesimal belt,
centred at 15 AU and 5 AU wide,  for two different values of the
planet eccentricity $e_{pl}$, i.e. 0.1 and 0.6, and for a warm and inclined population. 
The values of $P_i$ and
$U_c$ used in the collisional evolution code are taken from the previous computations.
The initial size distribution (red line), divided in a series of discrete bins, is truncated
at 10 and 500 km in diameters.
The erosion of the belt is
faster in the case labelled {\it 0.6w} (magenta circles) where $e_{pl}= 0.6$ and $e_{p0} = 0.5 \cdot e_f$,
 i.e. a warm belt. The initial cold belt with the same planet eccentricity ({\it 0.6c})  
(blue filled circles) differs only slightly from the {\it 0.6w} case. When the inclination 
of the warm belt is increased to $15^o$ the collisional evolution is reduced, as expected 
by the strong decrease of $P_i$ (see Fig.\ref{incli1}). The case with 
$e_{pl}= 0.1$, labelled {\it 0.1c}, illustrates the evolution of a cold belt perturbed 
by a low inclination planet and it is the less collisionally active case. }
\end{figure}

We have run four different models: two for cold belts  
with $e_{pl}= 0.1$ and $e_{pl}= 0.6$,
one for an 
initially warm belt where $e_{pl}= 0.6$ and  $e_{p0} = 0.5 \cdot e_f$, and the 
last case for $e_{pl}= 0.6$,  $e_{p0} = 0.5 \cdot e_f$  and $i=15^o$.  
The evolved planetesimal populations in the 
four different cases are shown in Fig. \ref{collev}.
When $e_{pl}= 0.6$, the erosion of the belt leads to a significant reduction 
in the number of planetesimals even at large sizes. Compared to the 
case with $e_{pl}= 0.1$,  the simulation with $e_{pl}= 0.6$ shows an
erosion rate about 10 times faster over 4 Gyr. 
A significant larger amount of dust is then 
expected to be produced when the eccentricity of the planet is 
higher, at least in the initial phases of evolution of the belt, as it could 
have been argued by the larger values of 
both  
$P_i$ and $U_c$.  
However, old belts may have been significantly depleted by the 
the initial fast collisional erosion and show at present a dust production rate 
comparable or even lower than 
that of young and less depleted belts around low eccentricity planets.

The collisional evolution of an initial warm belt with $e_{p0} = 0.5 \cdot e_f$  
and $e_{pl}= 0.6$ is only slightly faster compared to an initial cold belt 
($e_{p0} = 0$) and a similar dust production rate is expected. 
This shows that, in spite of an increase of the impact
velocity $U_c$ in the warm case, the two dynamical
configurations lead to a similar collisional evolution because
$P_i$  is almost equal. 
This is confirmed also by the inclined case where a lower 
value of $P_i$ leads to a
less eroded belt in 
spite of a higher collisional velocity (see Fig.\ref{incli1}). 
This suggests that the
intrinsic probability of collision is decisive in 
determining the erosion of a small body belt 
notwithstanding different values of impact velocity.

The previous modelling must be considered only as indicative 
since 
additional information are needed in particular about 
the structural strength and porosity of the planetesimals
since these physical aspects
are heavily involved in the computation of the
outcome of mutual collisions. 
In addition, this code is limited in the size range that it can cover 
and it does not possess spatial resolution. It cannot be compared with 
codes like DyCoSS and LIDT-DD \citep{theb2012,kral2013} describing the 
evolution of a debris disk both in time and space, including effects 
like the Poynting--Robertson
drag and creating a coupling between dynamics and collisions.
The only advantage of our simplified code is that it can predict on 
the long term (timescale of the order of Gyrs) how the dust production rate 
will decline and establish a potential correlation between the 
age of the star and the luminosity of the belt. It can also be used in a reverse 
approach allowing to determine the size distribution of the primordial belt 
from the present belt around an aged star, as it is usually done for 
the asteroid belt in the solar system. 

\section{The belt in HD 38529: comparison with $\epsilon$ Eridani}

\begin{figure}
  \includegraphics[width=\hsize]{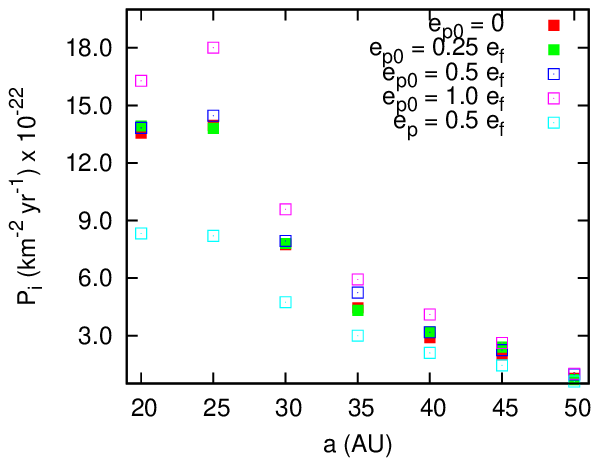}
  \includegraphics[width=\hsize]{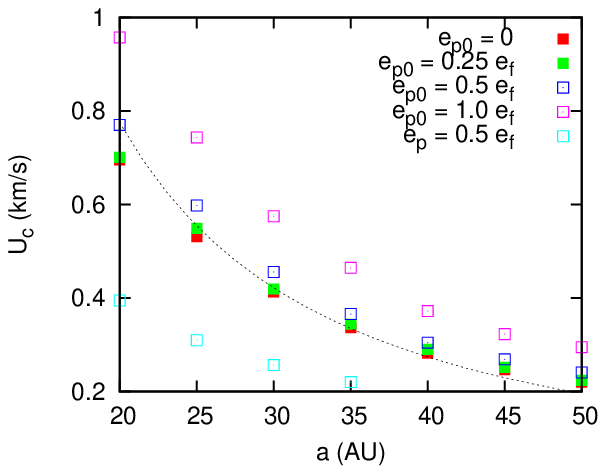}
\caption{\label{HD} $P_i$ (upper panel) and $U_c$ (lower panel) for the belt 
observed in the HD 38529 system. We consider three different models:
an initial cold belt 
with $e_p= e_f$ ($e_{p0} = 0$), an initial warm belt with different 
values of $e_{p0}$ and a collisionally damped belt with $e_p= 0.5 \cdot e_f$. 
}
\end{figure}

HD 38529 is a complex system where two massive planets have been found to orbit the 
central body, a 3.5 Gyr old G8 type star. 
Their masses are 0.8 and 12.2 $M_J$ and their orbital elements are $a_1 = 0.13$ AU 
, $a_2 = 3.74$ AU, $e_1 = 0.25$ and $e_2= 0.35$, respectively \citep{butler2006}. 
\cite{moro2007} found from Spitzer data an infrared excess that interpreted as due to 
the presence of a debris disk. From dynamical constraints they located the dust--producing 
long--lived 
planetesimals 
in three regions. A small inner ring orbiting between the two planets from 0.4 to 0.8 AU, 
a wider outer belt extending from 20 to 50 AU and an even outer one beyond 60 AU.  
The two outer belts are separated from a strong secular resonance located at about
55 AU.
We focus our study on the belt extending from 20 to 50 AU 
where \cite{moro2007} constrain the location of the dust--producing planetesimals. 
They in fact find that 
the debris disk is collision--dominated  implying that the planetesimals approximately share the 
same orbits as the emitting dust.  We compute for this system the intrinsic 
probability of collision and impact velocity at different radial distances within  
20 and 50 AU. 
In Fig. \ref{HD} we show $P_i$ and $U_c$ for this belt. We consider only the outer more 
massive planet as perturber and we neglect the influence of the smaller inner one.  Compared to the 
Kuiper Belt in the solar system, the intrinsic probability of collision appears 
to be larger in particular in the inner regions of the belt. According to \cite{oro2001}, 
$P_i$ in the Kuiper Belt ranges from 3.14 to 4.44 $\times 10^{-22}$ $yr^{-1}$
$km^{-2}$ while from Fig. \ref{HD} for 38529 $P_i$ is at least three times larger
in between 20 to 30 AU. On the other hand, the impact velocity in the HD 38529
debris disk appears to be smaller than that in the Kuiper belt whose
value is about 1.23--1.44 $km/s$. In short, in HD 38529 collisions are expected
to be more frequent but less energetic. 

\begin{figure}
  \includegraphics[width=\hsize]{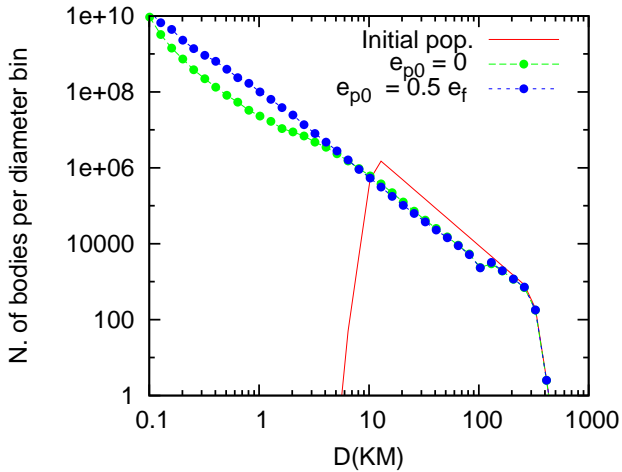}
\caption{\label{HD_COLL} Collisional evolution of a planetesimal belt
in the HD 38529. The erosion rate is much slower compared to that of 
the standard case shown in Fig. \ref{collev} due to the lower 
values of both $P_i$ and $U_c$.  The warm belt is slightly collisionally 
more active since a larger number of bodies are disrupted in 
between 10 and 40 km in diameter. These breakups produce smaller 
fragments which refill the population at small sizes.
}
\end{figure}

By using the values of $P_i$ and $U_c$ at the centre of the belt (35 AU) 
(see Fig. \ref{HD}) we run the 
collisional evolution code. In Fig. \ref{HD_COLL} the evolution with 
time of the size distribution of an
initial planetesimal belt equal to that used for Fig. \ref{collev} is 
shown. Compared to Fig. \ref{collev},  the shattering events are significantly 
reduced and only a limited amount of bodies smaller than 100 km in diameter
are catastrophically disrupted.  This must be ascribed to the lower value of 
$P_i$ which is about 30 times smaller for the HD 38529 belt compared 
to the $P_i$ of our standard case. 
The dust production rate is then expected to be significantly lower.
The difference observed between the cold and warm case at small
sizes is due to the slightly larger number of big bodies disrupted in the warm belt,
phenomenon 
which can barely be seen in Fig. \ref{collev} in between 10 and 40 km. The fragments 
produced in these breakups refill the size distribution at smaller diameters explaining the 
larger number of small bodies in the warm belt. 

Together with HD 38529, also $\epsilon$ Eridani has a debris disk identified by 
dust emission \citep{greaves98,mcgregor2015}. The belt is supposed to extend from 
approximately 53 to 80 AU from the central star and it is perturbed by a 
giant planet with a mass approximately 1.66 times the mass of Jupiter
and on an 
orbit with semimajor axis equal to 3.39 AU and eccentricity of 0.7 \citep{benedict06}.
This belt has approximately the same radial extension of that around 
HD 38529 (about 30 AU) but it is located farther out from the star. 
The mass in the dust estimated in the $\epsilon$ Eridani debris disk 
is about $2 \times 10^{-5} M_{\odot}$, according to \cite{li2003}.
This value is significantly higher compared to that inferred for 
HD 38529 by \cite{moro2007} of $1-5 \times 10^{-10}$ even if this 
was obtained for particles 
of 10 $\mu$m in size. A higher dust mass for $\epsilon$ Eridani can only partially 
be explained by the younger age of the system (about 800 Myr for $\epsilon$ Eridani 
vs. 3.28 Gyr for HD 38529). We have computed the intrinsic probability of 
collision and impact velocity for the putative planetesimal belt precursor 
of the debris disk in $\epsilon$ Eridani. At the centre of the belt around 
67 AU we find $P_i \sim 2 \times 10^{-23}$ $km^{-2} yr^{-1}$ and an 
average impact velocity of about $U_c \sim 0.1 km/s$ while 
for HD 38529 in the centre of the belt at 35 AU (see Fig. \ref{HD}) 
we have $P_i \sim 4.5 \times 10^{-22}$ $km^{-2} yr^{-1}$  
and $U_c \sim 0.35 km/s$ (assuming in both cases an initial cold belt).
The collisional activity is then expected to be
at least 20 times faster in HD 38529 respect to $\epsilon$ Eridani.  
However, as shown in Fig.\ref{HD_COLL}, the collisional evolution in
HD 38529 cannot have led to a significant erosion of the belt 
even after 3.28 Gyr. 
Why is then the dust in the debris disk around $\epsilon$ Eridani
more dense than that around HD 38529? The only possible explanation is 
that the planetesimal belt in $\epsilon$ Eridani is significantly 
more populated respect to that of HD 38529. The star of $\epsilon$ Eridani is 
less massive compared to that of HD 38529  (0.83 vs. 1.39 $M_{\odot}$) so it
is difficult to imagine that the protoplanetary disk around $\epsilon$ Eridani  
was much more dense than that around HD 38529. One possibility is that 
the dynamical history of HD 38529 has been more turbulent with 
extended periods of planet--planet scattering. This behaviour might have 
depleted the belt, as suggested by \citep{bonsor2013,marzari2014}, leaving 
a lighter belt that subsequently evolved under mutual collisions. This evolution 
might explain the present difference between the two systems.

\section{Discussion and conclusions}

The dust production rate in a debris disk, and then its brightness, strongly 
depends on the collisional evolution of the parent planetesimal belt. Since planetesimals are 
remnants of the planet formation process it is then reasonable to expect that one
or more planets formed in the system, notwithstanding any significant correlation between stars
with dust emission and the presence of known planets is still under scrutiny. 
The collisional evolution of planetesimals and the dust production rate can be determined 
once both
the physical structure of the bodies (in particular strength and porosity)  and 
two fundamental dynamical parameters, the intrinsic probability of collision $P_i$ and the 
mutual impact velocity $U_c$, are known. While the physical structure of planetesimals
depends on the dust accumulation process in the early phases of the circumstellar 
disk evolution, the values of the two parameters $P_i$ and $U_c$ are uniquely determined by the 
orbital architecture of the system. This in turn depends on  
planetesimal accumulation and planet formation but it is possibly influenced also by 
other evolutionary processes like  
planet migration by 
tidal interaction with 
the gaseous disk and phases of planet--planet scattering. This last mechanism is expected to 
have occurred in a significant fraction of systems due to the large number of exoplanets 
found in 
highly eccentric orbits. Planet--planet scattering not only excites eccentricity but it
also moves planets around during the chaotic period characterised by mutual 
close encounters between the planets prior to the ejection of one (or more) out of 
the system. 

A planetesimal belt can be affected by the dynamical behaviour of the planets  
and we can envision different scenarios in which planets and planetesimal
belts interact, evolve and finally coexist. 
We are interested in a scenario where a planet perturbs the belt from an inside
eccentric orbit. This configuration can be achieved 
via a smooth path where the planet, after its 
formation, evolved close to the belt without 
dramatic dynamical events and in a low eccentricity orbit. In this case the belt is 
expected to be initially 
cold before the onset of the secular perturbations of the planet. 
If instead the planet is involved in a period of planet--planet scattering, 
its present location close to the belt would be  
the outcome of a chaotic past and its eccentricity is expected to be high.  In this 
second scenario, the belt may have been affected by a period of planet--planet
scattering at different levels. It may have been spared by most of the chaotic 
evolution of the planets and, when one planet ends up in a close orbit, 
be  directly affected by its secular perturbations starting from a non--excited 
state (initial cold belt). In alternative, it may have been stirred up by the 
planets evolving on highly eccentric orbits induced by the mutual 
close encounters before a quiet state is established following the ejection of 
one or more bodies. In this case a warm belt with excited eccentricities and,
possibly, inclinations would interact with the planet which finally ends up in 
an orbit  
close to the belt at the end of the chaotic phase. 
Intermediate states can also be envisioned 
where the belt is warmed either by self--stirring induced by the formation of large embryos or 
by resonance sweeping by a migrating planet or other possible mechanisms 
before being perturbed by an incoming planet. 

In this paper we 
apply a semi-analytic method specifically designed to compute
both $P_i$ and $U_c$  for belts perturbed by a nearby planet. 
Its secular perturbations force a strong correlation between 
the eccentricity $e$ and perihelion longitude
$\varpi$ of the planetesimal orbits significantly affecting the values of both 
$P_i$ and $U_c$. 
The knowledge of these two dynamical parameters allow to 
model the collisional evolution of a belt and 
quantitatively predict its erosion 
over the star age. With suited collisional models these parameters can also 
be used to infer 
the early planetesimal population that led to the present belt 
whose properties are deduced by the related debris disk. From an observational point of view, they can 
be used to relate the star age to the brightness of the debris disk, taking of 
course into account the uncertainties on the initial planetesimal population,
its physical properties and the planet orbital parameters. 
It can also predict the frequency of large breakup events
in the belt which can lead to a sudden significant enhancement of the brightness of the           
related debris disk. The parameters we compute can also be used in more refined codes
predicting the luminosity of a debris disk for a given planetesimal belt.

We consider distinct
dynamical 
configurations in which the planet perturbs either an initial cold or warm belt. In the case of 
a warm belt, we also contemplate the case in which there is a significant inclination 
excitation. In addition, we compute the values of $P_i$ and $U_c$ for crowded belts 
where a significant collisional damping may occur. 
We focus on a standard case 
with a Jupiter--sized planet orbiting at about 5 AU from the star and a planetesimal
belt extending beyond 10 AU. We show that the secular perturbations of the 
planet and, in particular, the correlation between $e$ and $\varpi$ typical 
of these perturbations, cause a reduction of both $P_i$ and $U_c$  respect
to a non--correlated case. This last case may occur, in absence of 
a close perturbing planet,  if the planetesimal 
belt self--stirred for example via the formation of large planetary embryos, 
or if it was perturbed by planets roaming around during a chaotic
phase and, when finally a quiet state is reached, it is located far away 
from any potentially perturbing bodies.  This finding suggests that the presence 
of a perturbing planet not always leads to a higher dust production rate 
and that self--stirred planetesimal belts might give origin to brighter 
debris disks. We have tested different values of the planet eccentricity
finding higher values of both $P_i$ and $U_c$ in presence of more eccentric planets.  

We also explore as $P_i$ and $U_c$ change when a belt 
was warm before being secularly perturbed by a planet. 
In this case, the approximation $e_p = e_f$ cannot be adopted 
and the proper eccentricity, after the onset of the planet 
perturbations, can assume a range of values from $e_f - e_{p0}$ 
to $e_f + e_{p0}$ where $e_{p0}$ is the average eccentricity
of the warm planetesimal belt before the planet begins to perturb
it.  In our modelling we assume that $e_{p0}$ is a fixed fraction 
of the forced eccentricity, an arbitrary choice simply dictated by the need of 
avoiding a random selection of initial values. We do not consider 
scenarios  where $e_{p0}$  is initially larger than $e_f$.
For warm belts we find an increase in the collisional activity 
respect to initially cold belts but they are still less collisionally 
eroded compared to  
the 'self--stirred' belts, at least 
until $e_{p0} \leq e_f$.  A warm belt is then expected to 
be gradually depleted on shorter timescales. 
A dynamically excited belt may also 
have high inclinations. In this configuration an increase of the relative 
velocity and a decrease of the intrinsic probability of collision are 
found. However, the two effects do not compensate and 
the reduction of $P_i$ leads to a lower erosion rate compared 
to a small body belt with low inclination.  This proves the 
dominating role of $P_i$ in determining the collisional evolution of 
a minor body population. 

We have applied our formalism to model the case of the debris disk detected in 
HD 38529. The collisional activity in the putative belt hosting the 
leftover planetesimals is low compared to the standard case we have 
studied in this paper and after 
4 Gyr of collisional evolution only a low percentage of bodies smaller 
than 100 km are expected to have been disrupted by collisions. 
However, the planetesimal belt appears to be more collisionally active 
compared to that detected around $\epsilon$ Eridani. This finding is at odds 
with the observations of the debris disks around the two stars with 
that around $\epsilon$ Eridani being more dense respect to that around 
HD 38529. One possible explanation is that a turbulent dynamical past 
depleted the belt around HD 38529 reducing the number density of 
planetesimals. 

On the basis of the secular theory, we can envisage the following evolution of 
the brightness of a debris disk associated with an initially cold belt. At the beginning, when
the planet approaches the belt, the dust production rate will slowly increase 
during the progressive randomisation of the pericentre. When the $\varpi$s are 
still in phase, low impact velocities are expected but at subsequent times 
the progressively different values of pericentre longitudes will lead to more energetic collisions 
and a higher dust production rate. The debris disk will become brighter. When the 
full randomisation is reached, from then on the dominant mechanism will be the slow erosion 
of the planetesimal belt due to mutual collisions and the brightness of the 
associated debris disk will slowly decrease with time.  The rate of luminosity decrease can be 
estimated on the basis of $P_i$ and
$U_c$ which are the basic parameters to determine the collisional evolution of the belt. 

The situation appears slightly more complex for an initially warm belt. At the beginning,
just after the arrival of the planet, the pericentre evolution quickly leads 
pseudo--librators to cross the orbits of circulators causing a fast rise in the 
collisional rate and an increase in the luminosity of the associated debris disk.
However, at subsequent time this initial highly collisional state will develop 
into a dynamical configuration where the  
the secular perturbations are fully developed and the erosion with follow the 
same path as in a cold belt even if with different speed due to the 
higher values of $P_i$ and
$U_c$. 

In presence of a strong collisional damping, like in a very crowded belt,  the dynamics is 
dominated by the pseudo--libration of all planetesimals and the collisional activity is strongly 
reduced with values of both $P_i$ and $U_c$ significantly smaller respect
to both the cases of cold and warm non--damped belts. 

It is clear that, on the basis of the values of $P_i$ and
$U_c$ we can derive different relations between the brightness of a disk 
and the age of the star, since the erosion of the belt will be 
faster or slower depending on the orbital parameters of the planet 
and the dynamical state of the belt at the onset of the secular perturbations. 

In this paper we have focused on the perturbations of a single planet
on a planetesimal belt, 
but systems of multiple giant planets perturbing coexisting belts can be found as well. 
In this last case a more complex secular evolution is expected. 
 However, our algorithm for computing $P_i$ and 
$U_c$ can be easily applied to these systems once the 
the frequencies of perihelion precession of all planets 
are computed. If the 
eccentricities of the planets are high, the linear secular theory may be a rough 
approximation for computing the real frequencies and higher order theories 
may be needed  \citep{micht2004,
libert2005}.
This problem will be faced in a subsequent paper. 

Concerning the contribution from mean motion resonances to the 
collisional evolution of a  belt, it 
can be considered negligible in a configuration where a single planet 
perturbs a wide belt considering the tiny resonance width even 
for highly eccentric planets. 
Together with secular resonances, mean motion resonances 
may cause local differences in the brightness of the debris disk 
associated to the belt but 
they are not expected to cause significant differences in 
the overall collisional evolution of wide belts.  
For narrow belts a dedicated 
dynamical exploration may be necessary to evaluate the 
contribution of resonances. 

\section*{Acknowledgments}
We thank an anonymous referee for his/her useful comments and suggestions
that helped to significantly improve the paper.  

\bibliographystyle{aa}
\bibliography{biblio}

\bsp

\label{lastpage}

\end{document}